\documentclass[letterpaper, 10 pt, conference]{ieeetrans}
\IEEEoverridecommandlockouts

\usepackage[utf8]{inputenc}
\usepackage{amsmath}
\usepackage{mathrsfs}
\usepackage{amssymb}
\usepackage{color}
\usepackage[font=footnotesize]{caption}
\usepackage{cite}
\usepackage{comment}
\usepackage{float}
\usepackage{graphicx}
\usepackage{bernsteinStyle}
\usepackage{float}
\usepackage{tikz}
\usepackage{pgfplots}

\usetikzlibrary{shapes,arrows,calc,positioning}
\tikzstyle{bigblock} = [draw, fill=blue!20, rectangle, 
    minimum height=6em, minimum width=8em]
\tikzstyle{medblock} = [draw, fill=blue!20, rectangle, 
    minimum height=4em, minimum width=4em]    
\tikzstyle{mux} = [draw, fill=black!20, rectangle, 
    minimum height=5em, minimum width=0.1em]    
\tikzstyle{smallblock} = [draw, fill=blue!20, rectangle, 
    minimum height=2.5em, minimum width=4em]
\tikzstyle{sum} = [draw, fill=blue!20, circle, node distance=1cm]
\tikzstyle{signal} = [coordinate]
\tikzstyle{pinstyle} = [pin edge={to-,thin,black}]
\tikzstyle{block} = [draw, fill=blue!20, rectangle, 
    minimum height=3em, minimum width=6em]
\tikzstyle{blockS} = [draw, fill=blue!20, rectangle, 
    minimum height=3em, minimum width=4em]  
\tikzstyle{sum} = [draw, fill=blue!20, circle, node distance=1.5cm]
\tikzstyle{gain} = [draw, fill=blue!20, regular polygon, regular polygon sides = 3, node distance=1.25cm, shape border rotate = -90]
\tikzstyle{mult} = [draw, fill=blue!20, circle, inner sep=0pt, minimum size=0.2cm,]
\tikzstyle{saturation block} = [draw, fill=blue!20,
    	path picture={
    		\pgfpointdiff{\pgfpointanchor{path picture bounding box}{north east}}%
    		{\pgfpointanchor{path picture bounding box}{south west}}
    		\pgfgetlastxy\x\y
    		\tikzset{x=\x*.4, y=\y*.4}
    		\draw (-1,0) -- (1,0) (0,-1) -- (0,1); 
    		\draw (-1,-.7) -- (-.5,-.7) -- (.5,.7) -- (1,.7);
    	}]
\tikzstyle{sat atan} = [draw, fill=blue!20, 
        path picture={
          \pgfpointdiff{\pgfpointanchor{path picture bounding box}{north east}}%
            {\pgfpointanchor{path picture bounding box}{south west}}
          \pgfgetlastxy\x\y
          \tikzset{x=\x*.05, y=\y*0.3}
          \draw (-7.5,0) -- (7.5,0) (0,-1.2) -- (0,1.2); 
          \draw[scale = 1,domain=-7.5:7.5,smooth,variable=\x,black] plot ({\x},{(tanh((0.6*\x)))});
        }]
\tikzstyle{input} = [coordinate]
\tikzstyle{output} = [coordinate]
\newcounter{example}

\pgfplotsset{compat=1.14}
\title{A Discrete-Time, Time-Delayed Lur'e Model\\ with Biased Self-Excited Oscillations}
\author{\large Juan Paredes, Syed Aseem Ul Islam, Omran Kouba, and Dennis S. Bernstein% <-this % stops a space
\thanks{Juan Paredes, Syed Aseem Ul Islam, and Dennis S. Bernstein are with the Department of Aerospace Engineering, University of Michigan, Ann Arbor, MI, USA. {\tt\small \{jparedes, aseemisl, dsbaero\}@umich.edu}  Omran Kouba is with the Department of Mathematics in the Higher Institute of Applied Sciences and Technology, Damascus, Syria. }
}

\begin{document}

\maketitle

\begin{abstract}

Self-excited systems arise in many applications, such as biochemical systems, mechanical systems with fluid-structure interaction, and fuel-driven systems with combustion dynamics.  
This paper presents a Lur'e model that exhibits biased self-excited oscillations under constant inputs.
The model involves asymptotically stable linear dynamics, time delay, a washout filter, and a saturation nonlinearity.
For all sufficiently large scalings of the loop transfer function, these components cause divergence under small signal levels and decay under large signal amplitudes, thus producing an oscillatory response.
A bias-generation mechanism is used to specify the mean of the oscillation.
The main contribution of the paper is a detailed analysis of a discrete-time version of this model.
\end{abstract}

\section{Introduction}

A self-excited system has the property that the input is constant but the response is oscillatory.
Self-excited systems arise in numerous applications, such as biochemical systems, fluid-structure interaction, and combustion.  
The classical example of a self-excited system is the van der Pol oscillator, which has two states whose asymptotic response converges to a limit cycle.
A self-excited system, however, may have an arbitrary number of states and need not possess a limit cycle.
Overviews of self-excited systems are given in \cite{JENKINS2013167,Ding2010}, and applications to chemical and biochemical systems are discussed in \cite{chance,gray_scott_1990,goldbeter_berridge_1996}.
%
%Limit cycles that arise from systems of second order are discussed in \cite[pp. 287--291]{Andronov}.
%
%Nonlinear oscillations are discussed in \cite[pp. 393--458]{pain2005}.
%
Self-excited thermoacoustic oscillation in combustors is discussed in \cite{Dowling1997,awad_1986,chen_2016}.
Self-excited oscillations of a tropical ocean-atmosphere system are discussed in \cite{munnich1991}.
Fluid-structure interaction and its role in aircraft wing flutter is discussed in \cite{blevins,friedmann,coller,cesnik}. 
Wind-induced self-excited motion and its role in the Tacoma Bridge collapse is discussed in \cite{scott2001}.

Models of self-excited systems are typically derived in terms of the relevant physics of the application. 
From a systems perspective, the main interest is in understanding the features of the components of the system that give rise to self-sustained oscillations.
Understanding these mechanisms can illuminate the relevant physics in specific domains and provide unity across various domains.

A unifying model for self-excited systems is a feedback loop involving linear and nonlinear elements; systems of this type are called {\it Lur'e systems}. 
Lur'e systems have been widely studied in the classical literature on stability theory \cite{khalil3rd}.
%
%
%
%
%
%Papers about delayed velocity feedback for self-excited oscillation: 
Within the context of self-excited systems, Lur'e systems under various assumptions are considered in \cite{Ding2010,jian2004,Zanette_2017,Gusman_2016,CHATTERJEE20111860,gstan2007,Tomberg1989,chua1979,aguilar2009,hang2002}.
%
%Ding2010 mentions Lure?
%
%
%
Application to thermoacoustic oscillation in combustors is considered in \cite{savaresi2001}.
Self-oscillating discrete-time systems are considered in \cite{vrasvan98,amico2001,amico2004,amico2011}.

%Paper about nonlinear identification of combustor limit cycle behavior with Lur'e model: \cite{savaresi2001}

%Papers about conditions for self-oscillation in continuous time systems: \cite{gstan2007}, \cite{Tomberg1989}, \cite{chua1979}

Roughly speaking, self-excited oscillations arise from a combination of stabilizing and destabilizing effects.
Destabilization at small signal levels causes the response to grow from the vicinity of an equilibrium, whereas stabilization at large signal levels causes the response to decay from large signal levels.
In particular, negative damping at low signal levels and positive damping at high signal levels is the mechanism that gives rise to a limit cycle in the van der Pol oscillator \cite[pp. 103--107]{nayfeh2008}.
Note that, although systems with limit-cycle oscillations are self-excited, the converse need not be true since the response of a self-excited system may oscillate without the trajectory reaching a limit cycle.
Alternative mechanisms exist, however; for example, time delays are destabilizing, and Lur'e models with time delay have been extensively considered as models of self-excited systems \cite{minorskypaper}.

%Papers about delayed velocity feedback for self-excited oscillation: \cite{jian2004}, \cite{Zanette_2017}, \cite{Gusman_2016}, \cite{Ding2010}, \cite{CHATTERJEE20111860}

\begin{figure}[h]
    \centering
    \resizebox{0.8\columnwidth}{!}{%
    \begin{tikzpicture}[>={stealth'}, line width = 0.25mm]
    \node [input, name=input] {};
    \node [smallblock, rounded corners, right of=input, minimum height = 0.5cm, minimum width = 0.5cm] (beta) {$\beta$};
    \node [sum, right = 0.75cm of beta] (sum1) {};
    \node[draw = white] at (sum1.center) {$+$};
    \node [smallblock, rounded corners, right = 0.7cm of sum1, minimum height = 0.6cm, minimum width = 0.8cm] (system) {$G(s)$};
    
    \draw [->] (sum1) -- node[name=usys, above] {$v_\rmb$} (system);
    \node [output, right = 2.2cm of system] (output) {};
    \node [smallblock, rounded corners, below = 0.6cm of system, minimum height = 0.6cm, minimum width = 0.8cm](diff){$W(s)$};
    \node [smallblock, rounded corners, right = 0.5cm of diff, minimum height = 0.6cm, minimum width = 0.8cm] (delay) {$e^{-T_\rmd s}$};
    %\node [saturation block, left = 0.8cm of diff, minimum width=1.25cm, minimum height=1cm] (satq) {};
    %\node [sat atan, left = 0.7cm of diff, minimum width=1.25cm, minimum height=2.5em] (satq) {};
    \node [smallblock, rounded corners, left = 0.5cm of diff, minimum height = 0.6cm, minimum width = 0.8cm](satq){$\SN$};
    \node [mult, below = 0.1cm of beta, minimum size=0.35cm] (mult1) {};
    \node [draw = white] at (mult1.center) {$\times$};
    
    \draw [draw,->] (input) -- node [name=u]{} node [very near start, above] {$v$} (beta);
    \draw [->] (u.center) |- (mult1);
    %\draw [->] (beta) -- node [above] {$+$} (sum1);
    %\draw [->] (mult1) -| node [very near end, xshift = -0.25cm] {$+$} (sum1);
    \draw [->] (beta) -- (sum1);
    \draw [->] (mult1) -| (sum1);
    \draw [->] (satq) -|  
    node [near start, above] {$v_\rmf$} (mult1);
    \draw [->] (system) -- node [name=y, very near end]{} node [very near end, above] {$y$}(output);
    \draw [->] (y.center) |- (delay);
    \draw [->] (delay) -- node [above] {$y_\text{d}$} (diff);
    \draw [->] (diff) -- node [above] {$y_{\text{f}}$}(satq);
    \end{tikzpicture}
    }
	\caption{\footnotesize Continuous-time, time-delayed Lur'e  model with constant input $u$ and bias generation.}
    \label{CT_TDL_offset_blk_diag}
\end{figure}
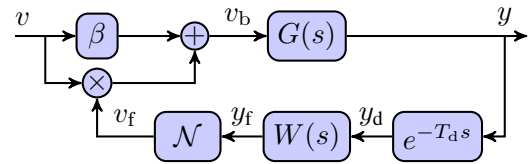

The present paper considers a time-delayed Lur'e (TDL) model that exhibits self-excited oscillations.
This model, which is illustrated in Figure  \ref{CT_TDL_offset_blk_diag}, incorporates the following components:
\begin{enumerate}
\item Asymptotically stable linear dynamics.
\item  Time delay.
\item  A washout (that is, highpass) filter.
\item  A continuous, bounded nonlinearity $\SN\colon\BBR\to\BBR$ that satisfies $\SN(0)=0$, is either nondecreasing or nonincreasing, and changes sign (positive to negative or vice versa) at the origin.
\item  A bias-generation mechanism, which produces an offset in the oscillatory response that depends on the value of the constant external input.
\end{enumerate}
%
%The components of this time-delay Lur'e model are motivated by the model given in \cite{paper}. 
%
%The bias-generation mechanism, which produces an offset in the oscillatory response that depends on the value of the constant external input, is a novel feature of the present paper.
%
A notable feature of this model is that self-oscillations are guaranteed to exist for  asymptotically stable dynamics that are not necessarily passive as in \cite{STAN2004LURELIN}.
%
%I am not sure about this.  Pls find a paper perhaps by Sepulchre that guarantees oscillations for PASSIVE linear dynamics
%DDDDDDD Got it. This paper is one of his previous ones. Here, the passive system is assumed to be linear.
%
We note that washout filters are used in \cite{abedwashout} to achieve stabilization, whereas, in the present paper, they are used to create self-oscillations.

%Figure  \ref{CT_TDL_offset_blk_diag} illustrates the components of this model.
%
For this time-delay Lur'e model, the time-delay provides the destabilization mechanism, while, under large signal levels, the saturation function yields a constant signal, which effectively breaks the loop, thus allowing the open-loop dynamics to stabilize the response.
This stabilization occurs at large amplitude.
In order to create an oscillatory response, the Lur'e model includes a washout filter, which removes the DC component of the delayed signal $y_\rmd$ and allows the saturation function to operate in its small-signal linear region.
A similar feature appears in \cite{jian2004,Zanette_2017,Gusman_2016,Ding2010,CHATTERJEE20111860} in the form of the numerator $s$ in $G$ for the case where $y$ represents velocity.
This combination of elements produces self-excited oscillations for all sufficiently large scalings of the asymptotically stable dynamics.
An additional feature of this model is the ability to produce oscillations with a bias, that is, an offset.
This is done by the bias-generation mechanism involving the scalar $\beta.$
Example \ref{ex_1_1} illustrates the response of the model in Figure  \ref{CT_TDL_offset_blk_diag}.
%
%
%

\begin{figure}[h!]
    \centering
    \includegraphics[width=\columnwidth]{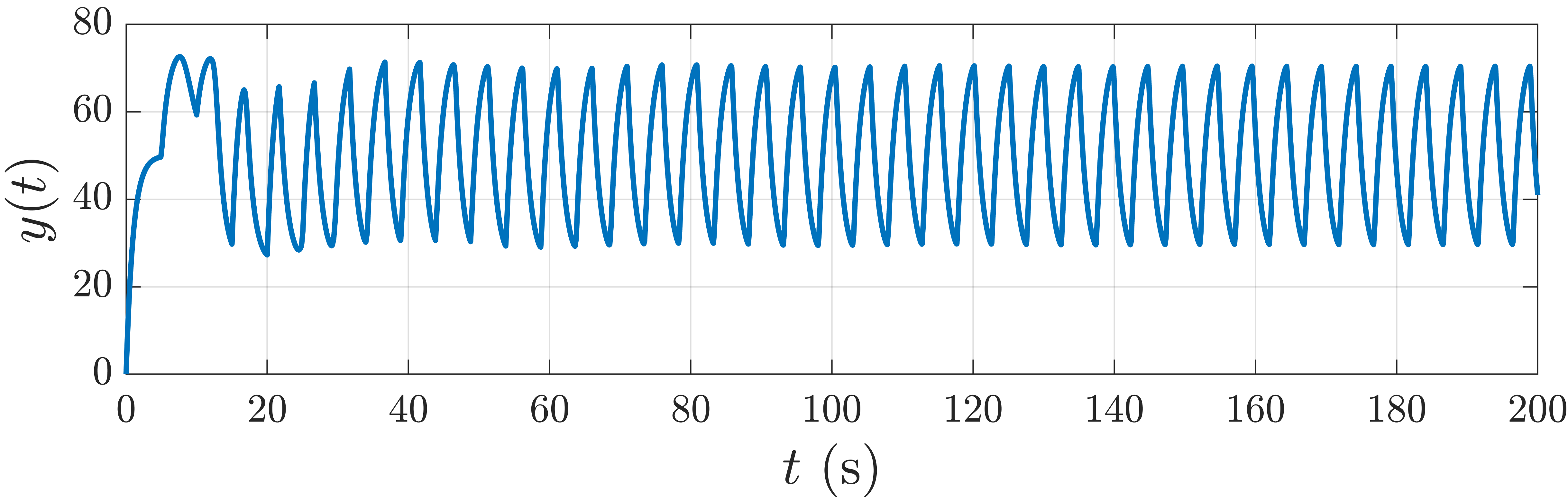}
    \caption{\footnotesize Example \ref{ex_1_1}  Self-excited oscillatory response of the continuous-time time-delay Lur'e model shown in Figure \ref{CT_TDL_offset_blk_diag}.  Note that the oscillation has nonzero bias due to the bias-generation mechanism.}
    \label{CTSimResults}
\end{figure}

\begin{figure}[h!]
    \centering
    \includegraphics[width=\linewidth]{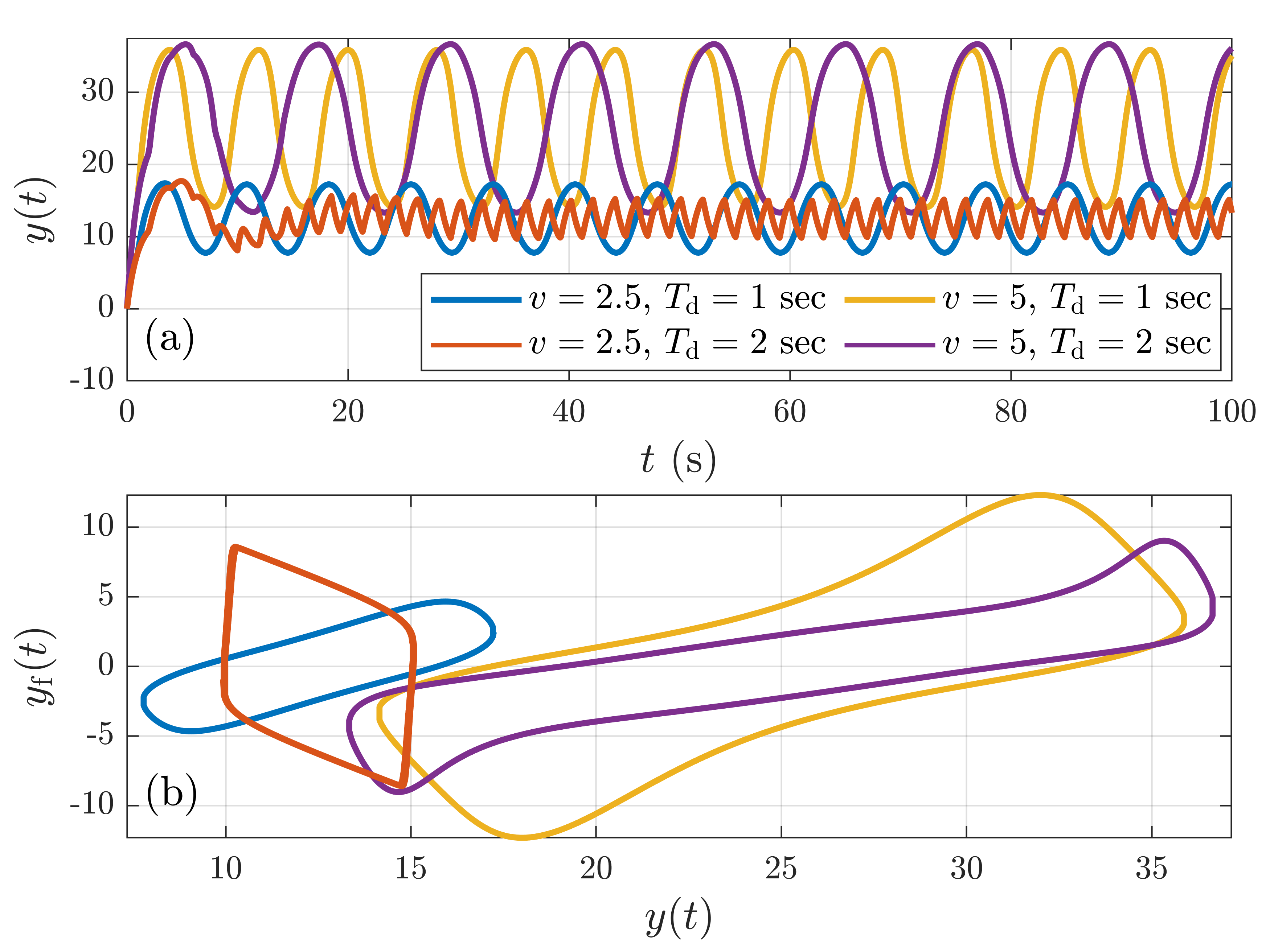} 
    \caption{\footnotesize Example \ref{ex_1_1}:  For $v = 2.5, 5,$ $\beta = 5,$ and $T_d = 1, 2$ s, (a) shows the response $y(t),$ and (b) shows $y(t)$ versus $y_{\rmf} (t)$ for $t>100$ s.   Note that the shapes of the  oscillations are distinct for different parameters under the same linear dynamics given by $G.$}
    \label{CTSimResults_1}
\end{figure}

\begin{figure}[h!]
    \centering
    \includegraphics[width=\linewidth]{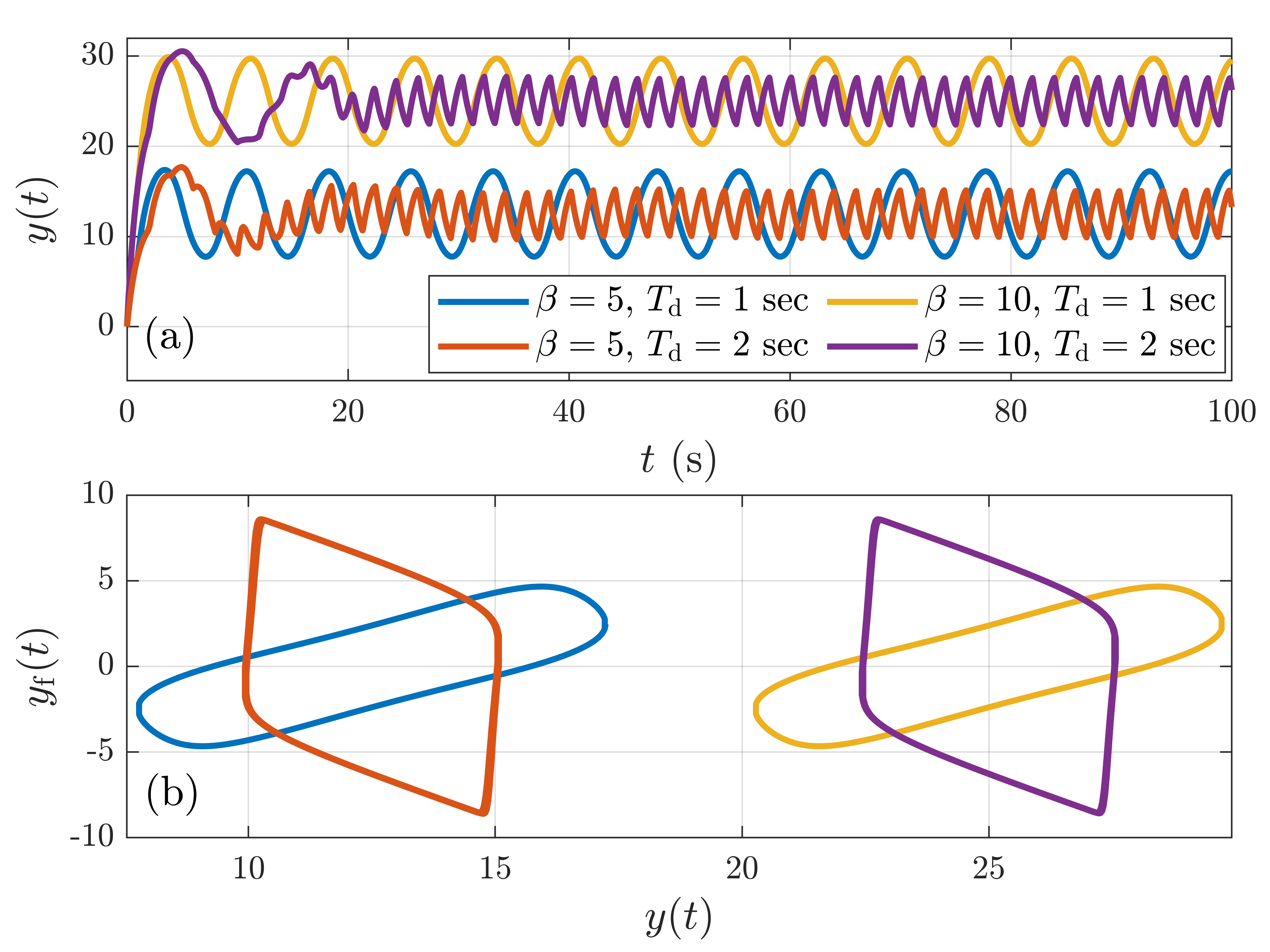}
    \caption{\footnotesize Example \ref{ex_1_1}:  For $v = 2.5,$ $\beta = 5, 10,$ and $T_d = 1, 2$ s, (a) shows the response $y(t),$ and (b) shows $y(t)$ versus $y_{\rmf} (t)$ for $t>100$ s. Note the effect of the value of $\beta$ on the output offset and how the shape of the oscillation is kept for different values of $\beta.$}
    \label{CTSimResults_2}
\end{figure}

%
%{\bf Example 1.1.} 
%
\begin{exam}
\label{ex_1_1}
%
%
%Let $G(s) = \frac{1}{s + 1},$ $W (s) = \frac{s}{0.001s + 1},$ and define $\SN$ by $v_{\rm f} = 2.5 \, {\rm tanh}(y_{\rm f}/2.5).$ 
%
Let $G(s) = \frac{1}{s + 1},$ $W (s) = \frac{s}{0.001s + 1},$ and 
$
    \SN(y_\rmf) = \tfrac{5}{2}  \tanh(\tfrac{2}{5}y_{\rm f}).$
    %
    % \SN(y_\rmf) = \tfrac{5}{2} \tanh(\tfrac{2}{5}y_{\rm f} - 2) + \tfrac{5}{2}\tanh(2).\label{Nnotodd}
%\end{gather}
%
% $\SN_1$ by $v_{\rm f} = 2.5 \, {\rm tanh}(y_{\rm f}/2.5)$ and $\SN_2$ by $v_{\rm f} = 2.5 \, {\rm tanh}((y_{\rm f} - 5)/2.5) + 2.4104.$
%
%Note that \eqref{Nodd} is  odd, but \eqref{Nnotodd} is not.
%
For  $T_{\rm d} = 5$ s, $\beta  = 5,$ and $v = 10,$ the response of the TDL model is shown in Figure \ref{CTSimResults}.
In particular, the output $y(t)$ converges to a periodic signal with bias $v \beta G(0) = 50.$
Next, the effect of $v,$ $\beta,$ and $T_\rmd$ on the oscillatory response of the model will be shown.
The response of the TDL model for $v = 2.5, 5,$ $\beta = 5,$ and $T_\rmd = 1, 2$ s is shown in Figure \ref{CTSimResults_1}, while the response for $v = 2.5,$ $\beta = 5, 10,$ and $T_\rmd = 1, 2$ s is shown in Figure \ref{CTSimResults_2}. 
Note that, for the same  transfer function $G$, different values of $v,$ $\beta,$ and $T_\rmd$ produce different waveforms for $y(t)$ and different phase portraits of $y_\rmf(t)$ versus $y(t).$
%
%Finally, define $\SN_1$ by $v_{\rm f} = 2.5 \, {\rm tanh}(y_{\rm f}/2.5)$ and $\SN_2$ by $v_{\rm f} = 2.5 \, {\rm tanh}((y_{\rm f} - 5)/2.5) + 2.4104.$
%
% Finally, it will be shown that $\SN$ does not need to be odd for the TDL model to yield an oscillatory response.
% %
% The response of the TDL model for $\SN = \SN_1, \SN_2,$ $T_{\rm d} = 5$ s, $\beta  = 5,$ and $v = 5,$ is shown in Figure \ref{CTSimResults_3}.
% %
% Note that whether nonlinearity $\SN$ is odd affects the waveforms produced for $y(t)$ for the same transfer function $G$ and values of $v,$ $\beta,$ and $T_\rmd.$
%
\hfill{\large$\diamond$}
\end{exam}

%
% \begin{figure}[h!]
%     \centering
%     \includegraphics[width=\linewidth]{figures/Section_1/s1_ex_1_3.png}
%     \caption{\footnotesize Example \ref{ex_1_1}:  For $v = 5,$ $\beta = 5,$ and $T_d = 5$ s, the figure shows the response $y(t)$ for the odd nonlinearity $\SN = \SN_1$ and for the non-odd nonlinearity $\SN = \SN_2.$  Note the effect of a non-odd nonlinearity on the output offset and the shape of the oscillations.}
%     \label{CTSimResults_3}
% \end{figure}

%
The analysis and examples in the paper focus on a discrete-time version of the time-delayed Lur'e model with the standard saturation function.
This setting simplifies the analysis of solutions as well as the numerical simulations.

The contents of the paper are as follows.
Section II considers a discrete-time linear feedback model and analyzes the range of values of $\alpha$ for which the closed-loop model is asymptotically stable.
Section III extends the problem in Section II by including a saturation nonlinearity.  This discrete-time Lur'e model is shown to have an asymptotically oscillatory response for sufficiently large values of the loop gain.
Section IV extends the Lur'e model to include a bias-generation mechanism.

%\subsection{Notation}

Preliminary results relating to the present paper appear in \cite{DTLACC}.   
Key differences between \cite{DTLACC} and the present paper include the following:  1) Lemma 2.1 and {\it v}) of Theorem 2.2 are not given in \cite{DTLACC}; 2)  due to limited space, no proofs are given in \cite{DTLACC}; and 3) the present paper includes several examples that do not appear in \cite{DTLACC}.

%The following definitions are needed.
%
Define 
$\BBZ\isdef\{\ldots, -1,0,1,\ldots\},$
$\BBN\isdef\{0,1,2,\ldots\},$ and
$\BBP\isdef\{1,2,\ldots\}.$
For all polynomials $p,$  ${\rm spr}(p)$ denotes the maximum magnitude of all elements of $\roots(p).$
For all nonzero $z=x+\jmath y\in\BBC$, where $x$ and $y$ are real,  $\arg z= \atann(y,x)\in(-\pi,\pi]$ denotes the principal angle of $z$.
%
%
%
%
%
%Let $P$ be a transfer function with no poles on the unit circle, and define $Z_P\isdef\{\theta\in[0,\pi]\colon P(e^{\jmath\theta})=0\}$.
%
Let $P \isdef N / D$ be a transfer function with no zeros on the unit circle,
where $N$ and $D$ are coprime, $m \isdef {\rm deg} N$ and $n \isdef {\rm deg} D.$ 
%
%and define $Z_P\isdef\{\theta\in[0,\pi]\colon N(e^{\jmath\theta})=0\}$.
%
For all $\theta \in [0,\pi],$  writing $P(z) = \frac{N(z)}{D(z)} = \frac{K \prod_{i=1}^{m} (z - z_i)}{\prod_{j=1}^{n} (z - p_i)},$ where $K$ is a nonzero real number, $\angle P(e^{\jmath \theta})\in\BBR$ denotes the unwrapped phase angle of $P$ evaluated at $\theta \in (-\pi,\pi],$ such that 
\begin{equation*}
    \angle P(e^{\jmath \theta}) \isdef \sum_{i=1}^{m} \arg(e^{\jmath \theta} - z_i) - \sum_{j=1}^{n}\arg(e^{\jmath \theta} - p_i).
\end{equation*}
Unlike $\theta\mapsto\arg P(e^{\jmath \theta}),$ which may be discontinuous on $[0,\pi]$, the function $\theta\mapsto\angle P(e^{\jmath \theta})$ is C$^1$ on $[0,\pi]$.
In addition, for all $\theta \in [0,\pi],$ there exists $r_\theta\in\BBZ$ such that $\angle P(e^{\jmath \theta}) = \arg P(e^{\jmath \theta}) + 2 \pi r_\theta.$

\section{Time-Delayed Linear Feedback model}

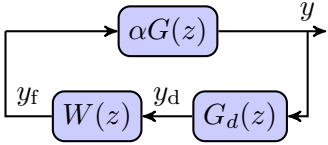
\begin{figure}[h]
    \centering
    %\vspace{1em}
     \resizebox{0.5\columnwidth}{!}{%
    \begin{tikzpicture}[>={stealth'}, line width = 0.25mm]
    \node [input, name=input] {};
    \node [smallblock,  rounded corners, right = 1.35cm of input, minimum height = 0.6cm, minimum width = 0.8cm] (system) {$\alpha  G(z)$};
    \draw [->] (input) -- node[name=usys] {} (system);
    \node [output, right = 1.3cm of system] (output) {};
    \node [smallblock, rounded corners, below left = 0.4cm and -0.3cm of system, minimum height = 0.6cm, minimum width = 0.8cm] (diff) {$W(z)$};
    \node [smallblock, rounded corners, right = 0.6cm of diff, minimum height=2.5em, minimum height = 0.6cm, minimum width = 0.8cm] (delay) {$G_d(z)$};
    \draw [-] (diff.west) -| node [near start, above] {$y_{\rm f}$} (input);
    \draw [->] (system) -- node [name=y,near end, xshift = 0.1cm]{} node [near end, above, xshift = 0.1cm] {$y$}(output);
    \draw [->] (y.center) |- (delay);
    \draw [->] (delay) -- node [above] {$y_\rmd$} (diff);
    \end{tikzpicture}
    }
    \caption{\footnotesize Discrete-time time-delayed linear feedback system.}
    \label{DT_TDL_woSat_blk_diag}
\end{figure}

In this section we consider the discrete-time, time-delayed Lur'e model shown in Figure \ref{DT_TDL_woSat_blk_diag}, where $\alpha \in \BBR,$ $G$ is a  strictly proper asymptotically stable SISO transfer function with no zeros on the unit circle, $G_d(z) \isdef 1/z^d$ is a $d$-step delay, where $d\in\BBN,$ and  $W(z) \isdef (z-1)/z$ is a washout (that is, highpass) filter.
Let $G=N/D$, where the polynomials $N$ and $D$ are coprime, $D$ is monic, $n\isdef \deg D$, and $m\isdef \deg N.$

Let $(A,B,C,0)$ be a minimal realization of $G$ whose internal state at step $k$ is $x_k\in\BBR^n.$
Furthermore, consider the realization $(N_d,e_{d,d},e_{1,d}^\rmT,0)$ of  $G_d$ with internal state $x_{\rmd,k}\in\BBR^d$, where $N_d$ is the standard $d\times d$ nilpotent matrix and $e_{i,d}$ is the $i$th column of the $d\times d$ identity matrix $I_d.$
Finally, let $(0,1,-1,1)$ be a realization of $W$ with internal state $x_{\rmf,k}\in\BBR,$ and let $\alpha$ be a real number that scales $G.$
%
%
%
%Now, let $(A_{\rm LF},0,C_{\rm LF})$ be a realization of t
%
Then, the discrete-time, time-delayed linear feedback model shown in Figure \ref{DT_TDL_woSat_blk_diag} has the closed-loop dynamics
\begin{align}
     \left[ \arraycolsep=1.1pt\def\arraystretch{1.2} \begin{array}{c} x_{k+1}\\ x_{\rmd,k+1}\\ x_{\rmf,k+1} \end{array} \right]
    =  \left[ \arraycolsep=1.6pt\def\arraystretch{1.2} \begin{array}{ccc} A & \alpha B e_{1,d}^\rmT & -\alpha B\\ e_{d,d}C & N_d & 0\\ 0 & e_{1,d}^\rmT   & 0 \end{array} \right]
     \left[ \arraycolsep=1.1pt\def\arraystretch{1.2} \begin{array}{c} x_{k}\\ x_{\rmd,k}\\ x_{\rmf,k} \end{array} \right] ,
\end{align}
with output    
\begin{align}
    y_{k} = \matl{ccc} C & 0 & 0 \matr
    \matl{c} x_{k}\\ x_{\rmd,k}\\ x_{\rmf,k}\matr
    \label{TDLeqnlin}
\end{align}
and internal signals 
\begin{align}
y_{\rmd, k} &= e_{1,d}^\rmT x_{\rmd, k},\label{ydk}\\
y_{{\rm f}, k} &= -x_{\rmf, k} + y_{\rmd, k}.\label{yfk}
\end{align}
For all $d\in\BBN $ and $\alpha\in\BBR$, define 
\begin{equation}
    L_{d,\alpha}(z) \isdef \alpha G(z) W (z) G_d (z) 
    =\frac{\alpha(z-1)N(z)}{z^{d+1}D(z)}.
\end{equation}
Furthermore, for all $d\in\BBN,$ define
$L_d \isdef L_{d,1} = GW G_d.$
Finally, for all $d\in\BBN $ and $\alpha\in\BBR,$ note that the closed-loop transfer function of the time-delayed linear feedback model
%from $v$ to $y$
is given by
\begin{align}
   %\tilde G_{y} =
   \frac{L_{d,\alpha}}{1-L_{d,\alpha}} = \frac{\alpha(z-1)N(z)}{p_{d,\alpha}(z)},
\end{align}
where
\begin{equation}
    p_{d,\alpha}(z)\isdef z^{d+1} D(z) - \alpha (z-1)N(z).
\end{equation}
% Finally, the characteristic polynomial $p_{d,\alpha}$ of the closed-loop model is given by
% %
% \begin{align}
% p_{d,\alpha}(z)\isdef z^{d+1} D(z) - \alpha (z-1)N(z).
% \end{align}
% %
Note that, for all $\alpha\in\BBR,$ 1 is not a root of $p_{d,\alpha}$.

The following lemma is needed for the proof of Theorem \ref{theorem_1}.
 
\begin{lem} \label{lemmaPQ}
Let $p$ and $q$ be monic polynomials with real coefficients,
assume that $\deg q<\deg p$, 
assume that all of the roots of $p$ are in the open unit disk, and, for all $\alpha\in\BBR,$ define $p_\alpha\isdef p+\alpha q.$
Then, there exist $\alpha_\rmn<0,$ $\alpha_\rmp>0,$ and $\delta>0$ such that
$\spr(p_{\alpha_\rmn})=\spr(p_{\alpha_\rmp})=1$,
for all $\alpha\in(\alpha_\rmn,\alpha_\rmn+\delta)\cup(\alpha_\rmp-\delta,\alpha_\rmp),$ $\spr(p_\alpha)<1,$
and, for all $\alpha\in(\alpha_\rmn-\delta,\alpha_\rmn)\cup(\alpha_\rmp,\alpha_\rmp+\delta),$ $\spr(p_\alpha)>1.$ 
\end{lem}

\textbf{Proof.}
Let $k$ be the smallest positive integer such that
\begin{equation}
	h(x)\isdef x^k[p(x)q(1/x)-q(x)p(1/x)]\nn
\end{equation}
is a polynomial, and define
\begin{equation*}
	\mathcal{Z}\isdef\{z\in\BBC\colon |z|=1,h(z)=0,\text{ and }q(z)\ne 0\}.
\end{equation*}
Note that $\mathcal{Z}$ has at  most $\deg h$ elements.
Furthermore, since $h(z)=0$ for all $z\in \mathcal{Z}$ and $p$ and $q$ have real coefficients, it follows that, for all $z\in \mathcal{Z},$ 
\begin{equation*}
    \overline{\left(\frac{p(z)}{q(z)}\right)}=\frac{p(\bar{z})}{q(\bar{z})}=\frac{p(1/z)}{q(1/z)}=	\frac{p(z)}{q(z)},
\end{equation*}
which implies that, for all $z\in \mathcal{Z},$  $p(z)/q(z)\in \mathbb{R}$.
%
% Now, if $z\in \mathcal{Z}$ then we conclude from $h(z)=0$ and the fact that $p$ and $q$ have real coefficients, that
% \begin{equation}
% \overline{	\left(\frac{p(z)}{q(z)}\right)}=\frac{p(\bar{z})}{q(\bar{z})}=\frac{p(1/z)}{q(1/z)}=	\frac{p(z)}{q(z)}.
% \end{equation}
%
%So $p(z)/q(z)\in \mathbb{R}$. 
%

Next, define $\mathcal{A}\isdef \{-p(z)/q(z)\colon z\in\mathcal{Z}\},$
and, in the case where $\mathcal{A}$ is not empty, let $\mathcal{A}=\{\alpha_1,\ldots,\alpha_m\},$ where $m\leq \deg h$ and $\alpha_1 < \cdots < \alpha_m$. 
Note that, since, for all $z\in\mathcal{Z},$ $p(z)\ne0,$ it follows that 
$0 \notin \mathcal{A}.$ 
Now, let $\alpha$ be a real number that is not contained in $\mathcal{A}$,
and suppose that $\spr(p_\alpha)=1.$
Then, there exists $z_\alpha\in\BBC$ such that $p_\alpha(z_\alpha) = 0$ and $|z_\alpha|=1.$
%
%If $\spr(p_\alpha)=1$ this means that there exists a zero $z_\alpha$ of $p_\alpha$ such that $|z_\alpha|=1$. %
%

To show that $q(z_\alpha)\ne0,$ suppose that $q(z_\alpha)=0.$
Then, since $p_\alpha(z_\alpha) = 0,$ it follows that $p(z_\alpha)=0,$ which, since all of the roots of $p$ are in the open unit disk, implies that $\spr(p_\alpha)<1,$ which is a contradiction.
Hence, $q(z_\alpha)\ne0.$
%
%Clearly, if $q(z_\alpha)\ne0$ because this would imply that $p(z_\alpha)=0$ and this is absurd.
%

Next, to show that $\spr(p_\alpha)\ne1,$ note that
$p_\alpha(z_\alpha) = 0$ implies 
\begin{align}
    0 &= \overline{p_\alpha(z_\alpha)}\nn\\ &=\overline{p(z_\alpha)+\alpha q(z_\alpha)}\nn\\
    &= \overline{p(z_\alpha)}+\alpha \overline{q(z_\alpha)}\nn\\
    &= {p(\overline{z_\alpha})}+\alpha q(\overline{z_\alpha})\nn\\
    &= p(1/z_\alpha)+\alpha q(1/z_\alpha). \label{pqalphaConjEq}
\end{align}
Since, in addition,
$
    \alpha = -p(z_\alpha)/q(z_\alpha),
$
it follows from 
\eqref{pqalphaConjEq} that
\begin{equation}
    p(1/z_\alpha) - (p(z_\alpha)/q(z_\alpha)) q(1/z_\alpha)=0.\label{qpzalph}
\end{equation}
Now, multiplying both sides of \eqref{qpzalph} by $q(z_\alpha)$ implies
\begin{equation*}
    q(z_\alpha) p(1/z_\alpha) - p(z_\alpha) q(1/z_\alpha) = -h(z_\alpha)/z_\alpha^k = 0,
\end{equation*}
and thus $h(z_\alpha)=0$.
%
%Further, taking conjugates we conclude from $p(z_\alpha)+\alpha q(z_\alpha)=0$ that we have also $p\left(\frac{1}{z_\alpha}\right)+\alpha \left(\frac{1}{z_\alpha}\right)=0$.
%
%Eliminating, $\alpha$ we conclude that $h(z_\alpha)=0$.
%
Hence, $z_\alpha\in\mathcal{Z}$, and thus $\alpha=-p(z_\alpha)/q(z_\alpha)\in\mathcal{A},$ which is a contradiction.
%
%Therefore,  $\spr(p_\alpha)\ne 1.$
%
%We conclude that $\spr(p_\alpha)\ne 1$ if and only if $\alpha\notin\{\alpha_1,\ldots,\alpha_m\}$.
%
Therefore, for all $\hat\alpha\notin\mathcal{A},$ $\spr(p_{\hat\alpha})\ne 1.$

Next, let $j \in \{0,1,\ldots,m\},$ and define $I_j\isdef(\alpha_j,\alpha_{j+1}),$ where $\alpha_0\isdef-\infty$ and $\alpha_{m+1}\isdef\infty$.
For all $\alpha \in I_j,$ it follows from the continuity of $\alpha\mapsto \spr(p_\alpha)$ that either $\spr(p_\alpha)<1$ or $\spr(p_\alpha)>1.$
Next, %we show that, for all $\alpha\in I_0\cup	I_m,$ $\spr(p_\alpha)>1.$
%
%Next, we show that $\spr(p_\alpha)>1$ if $\alpha\in I_0\cup	I_m$.
%
write
\begin{equation*}
	p(x)=a_nx^n+\cdots+a_0,\quad
	q(x)=b_dx^d+\cdots+b_0
\end{equation*}
such that $b_d\ne0$, $a_n\ne0,$ and $d<n$,
and let $z_{1,\alpha},\ldots,z_{n,\alpha}$ be the roots of $p_\alpha.$
%
%Rearranging $p_\alpha,$ 
%
%
%
%
%it follows from Vieta's formulas \cite{xxxx} that
%
% Suppose that
% 	\begin{equation}
% 		p(x)=a_nx^n+\ldots+a_0,\quad\textrm{and}\quad
% 		q(x)=b_dx^d+\ldots+b_0
% 	\end{equation}
% with $b_d,a_n\ne0$ and $d<n$. If $z_1^{(\alpha)},\ldots,z_n^{(\alpha)}$ are the roots of $p_\alpha$ (each repeated according to its multiplicity) then
%
%
%
Then, %it follows from Vieta's formulae that 
the coefficient of $x^d$ in $p_\alpha$ is related to the roots of $p_\alpha$ by
\begin{equation}
	a_d+\alpha b_d=a_n(-1)^{n-d} \sum\prod_{j\in B}z_{j,\alpha},\label{vieta}
\end{equation}
%
%
%_{\stackrel{B\subset\{1,\ldots,n\}}{|B|=n-d}}
%
%
where the sum is taken over all $\binom{n}{n-d}$ subsets $B$ of $\{1,\ldots,n\}$ with $n-d$ elements.
It thus follows from \eqref{vieta} that
\begin{equation*}
	|a_d+\alpha b_d|\le|a_n|\binom{n}{n-d}\spr(p_\alpha)^{n-d},
\end{equation*}
which implies that
\begin{equation*}
	\lim_{\alpha\to-\infty}\spr(p_\alpha) = \lim_{\alpha\to\infty}\spr(p_\alpha)=\infty.
\end{equation*}
Hence, for all $\alpha \in I_0 \cup I_m,$ $\spr(p_\alpha)>1.$
%
%In particular, we must have $\spr(p_\alpha)>1$ on the intervals $I_0$ and $I_m$.

Next, since $\spr(p_0)=\spr(p)<1,$ $0\notin\mathcal{A},$ and, for all $\alpha \in I_0 \cup I_m,$ $\spr(p_\alpha)>1$, it follows that there exists a unique $j_0\in\{1,\ldots,m-1\}$ such that $0\in I_{j_0}.$
Hence, for all $\alpha\in I_{j_0},$ $\spr(p_\alpha)<1.$
%
%On the other hand we know that $\spr(p_0)<1$ so there exists a unique$j_0\in\{1,2,m-1\}$ such that $0\in I_{j_0}$, so, we have $\spr(p_\alpha)<1$ for $\alpha\in I_{j_0}$.
%
%
 % 
%
Now, define 
\begin{align*}
j_\rmn&\isdef\min\{j\in\{1,\ldots,m-1\}\colon\spr(p_\alpha)<1 \text{ for all }\alpha\in I_{j}\},\\
j_\rmp&\isdef\max\{j\in\{1,\ldots,m-1\}\colon\spr(p_\alpha)<1 \text{ for all }\alpha\in I_{j}\}.
\end{align*}
Then, for all $\alpha\in I_{j_\rmn}\cup I_{j_\rmp},$ $\spr(p_\alpha)<1$ and, for all $\alpha\in I_{j_\rmn - 1}\cup I_{j_\rmp + 1},$  $\spr(p_\alpha)>1,$
and thus it follows from the continuity of $\spr$ and the intermediate value theorem that $\spr(p_{\alpha_{j_\rmn}}) = \spr(p_{\alpha_{j_\rmp + 1}}) = 1.$
Furthermore, since $j_\rmn \leq j_0 \le j_\rmp,$ it follows that $\alpha_{j_\rmn} < 0$ and $\alpha_{j_\rmp + 1} > 0.$
Hence, defining $\alpha_\rmn\isdef\alpha_{j_\rmn}$ and $\alpha_\rmp\isdef\alpha_{j_\rmp + 1},$
which, as an aside, shows that $\mathcal{A}$ has at least two elements,
it follows that $\alpha_\rmn<0,$ $\alpha_\rmp>0,$ and $\spr(p_{\alpha_\rmn}) = \spr(p_{\alpha_\rmp}) = 1,$ and, furthermore,
there exists $\delta>0$ such that,
for all $\alpha \in (\alpha_\rmn,\alpha_\rmn+\delta)\cup(\alpha_\rmp-\delta,\alpha_\rmp),$ $\spr(p_\alpha)<1,$
and,
for all $\alpha \in (\alpha_\rmn-\delta,\alpha_\rmn)\cup(\alpha_\rmp,\alpha_\rmp+\delta),$ $\spr(p_\alpha)>1,$
which completes the proof.
\hfill$\square$

%
%which concludes the proof.
%
% Then $\spr(p_\alpha)<1$ for $\alpha\in I_{j_1}$ and
% $\spr(p_\alpha)>1$ for  $\alpha\in I_{j_1+1}$. This achieves the proof of the lemma,
% with $\alpha_0=\alpha_{j_1+1}$ and $\delta=\min(\alpha_{j_1+1}-\alpha_{j_1},
% \alpha_{j_1+2}-\alpha_{j_1+1})$.

The following result shows that, for sufficiently large values of the delay $d,$ the linear closed-loop system is not asymptotically stable outside of a bounded interval of values of $\alpha.$
This result also shows that, for asymptotically large $d,$ this range of values of $\alpha$ is finite and symmetric.

%\textbf{Theorem 1.} 
\begin{theo}
\label{theorem_1}
The following statements hold:
%
%%%%%%%%%%%%%%%%%%PREVIOUS THEOREM
%\begin{enumerate}
%
%
% %
% %
% \begin{align} 
%     \frac{\rmd}{\rmd\theta}\angle L(e^{\jmath\theta}) < 0.\label{angLTheta_dtheta_leq0}
% \end{align}
% %
% %
% %
%     \item  There exist $\alpha_{d,\rml}<0$ and $\alpha_{d,\rmu} > 0$ such that $p_{d,\alpha}$ is asymptotically stable if and only if $\alpha\in(\alpha_{d,\rml},\alpha_{d,\rmu})$. 
% %
%%%%%%%%%%%%%%%%%%%%%NEW THEOREM
%
%
\begin{enumerate}
%
% \item For all $d \in \BBN,$ there exist $\alpha_{d,0},\alpha_d,\alpha_{d,1}>0$ such that $\alpha_{d,0} < \alpha_d< \alpha_{d,1},$ 
% %
% and
% ${\rm spr}(p_{d, \alpha_{d,0}}) < {\rm spr}(p_{d, \alpha_d})=1 < {\rm spr}(p_{d, \alpha_{d,1}}),$ and, for all $\alpha \in(\alpha_{d,0},\alpha_{d,1}),$ ${\rm spr}(p_{d, \alpha_{d,0}}) < {\rm spr}(p_{d, \alpha}) < {\rm spr}(p_{d, \alpha_{d,1}}).$
%
\item For all $d \in \BBN,$ there exist $\alpha_{d,0},\alpha_d,\alpha_{d,1}>0$ such that $\alpha_{d,0} < \alpha_d< \alpha_{d,1},$ 
${\rm spr}(p_{d, \alpha_d})=1,$
for all $\alpha \in (\alpha_{d,0}, \alpha_d),$ ${\rm spr}(p_{d, \alpha})<1,$
and,
for all $\alpha \in (\alpha_d, \alpha_{d,1}),$ ${\rm spr}(p_{d, \alpha})>1.$
% 
% \item For all $d \in \BBN,$ there exist $\alpha_{d,0},\alpha_d,\alpha_{d,1}<0$ such that $\alpha_{d,1} < \alpha_d< \alpha_{d,0},$
% %
% ${\rm spr}(p_{d, \alpha_{d,0}}) < {\rm spr}(p_{d, \alpha_d})=1 < {\rm spr}(p_{d, \alpha_{d,1}}),$ and, for all $\alpha\in(\alpha_{d,1},\alpha_{d,0}),$ ${\rm spr}(p_{d, \alpha_{d,0}}) < {\rm spr}(p_{d, \alpha}) < {\rm spr}(p_{d, \alpha_{d,1}}).$
%
\item For all $d \in \BBN,$ there exist $\alpha_{d,0},\alpha_d,\alpha_{d,1}<0$ such that $\alpha_{d,1} < \alpha_d< \alpha_{d,0},$
${\rm spr}(p_{d, \alpha_d})=1,$
for all $\alpha \in (\alpha_d, \alpha_{d,0}),$ ${\rm spr}(p_{d, \alpha})<1,$
and,
for all $\alpha \in (\alpha_{d,1}, \alpha_d),$ ${\rm spr}(p_{d, \alpha})>1.$
\end{enumerate}
Furthermore, there exists $\bar{d}\in\BBN$ such that the following statements hold:
\begin{enumerate}
\item[{\it iii})]  For all $d>\bar{d}$ and $\theta\in(0,\pi]$, $L_d(e^{\jmath\theta})\ne0$ and
 \begin{align} 
     \frac{\rmd}{\rmd\theta}\angle L_d(e^{\jmath\theta}) < 0.\label{angLTheta_dtheta_leq0}
 \end{align}
\item[{\it iv})]  For all $d>\bar{d},$ there exist $\alpha_{d,\rml}<0$ and $\alpha_{d,\rmu} > 0$ such that $p_{d,\alpha}$ is asymptotically stable if and only if $\alpha\in(\alpha_{d,\rml},\alpha_{d,\rmu}),$
and $p_{d,\alpha}$ is not asymptotically stable if and only if $\alpha\in(-\infty,\alpha_{d,\rml}]\cup[\alpha_{d,\rmu},\infty).$

\item[{\it v})]  Define
\begin{equation}
\alpha_{\infty}\isdef \min_{\theta\in(0,\pi]}\left|\frac{D(e^{\jmath\theta})}{(e^{\jmath\theta}-1)N(e^{\jmath\theta})}\right|. \label{alphaInfEq}
\end{equation}
Then, $\alpha_\infty>0,$ for all $d>\bar{d},$ $\alpha_\infty \leq \min\{-\alpha_{d,\rml}, \alpha_{d,\rmu}\},$ and
%
% where
% %
% \begin{equation}
% \alpha_{\infty}\isdef \min_{\theta\in(0,\pi]\backslash Z}\left|\frac{D(e^{\jmath\theta})}{(e^{\jmath\theta}-1)N(e^{\jmath\theta})}\right| > 0, \label{alphaInfEq}
% \end{equation}
% 
%
\begin{equation}
    \lim_{d\to\infty} -\alpha_{d,\rml} = \lim_{d\to\infty} \alpha_{d,\rmu}=\alpha_{\infty}. \label{alphainf}
\end{equation}
%
% \begin{align}
%     \alpha_{\infty}\isdef \lim_{d\to\infty} -\alpha_{d,\rml} = \lim_{d\to\infty} \alpha_{d,u}=\min_{\theta\in(0,\pi]\backslash Z}\left|\frac{D(e^{\jmath\theta})}{(e^{\jmath\theta}-1)N(e^{\jmath\theta})}\right|  > 0, \label{alphainf}
% \end{align}
\end{enumerate}
\end{theo}

\medskip

{\bf Proof.}  
{\it i}) and {\it ii}) follow from Lemma \ref{lemmaPQ}.
%
% To prove {\it i}) and {\it ii}),
% %
% let $d\in\BBN$ and note that, for all $\alpha\in\BBR,$ $p_{d,\alpha} = p_{d,0} - \alpha q,$ where $p_{d,0}(z) = z^{d+1}D(z)$ and $q(z) \isdef (z-1)N(z).$
% %
% Note that ${\rm spr}(p_{d, 0}) < 1.$ 
% %
% Furthermore, since $\deg q = m+1 < n+d+1 = \deg p_{d,0},$ it follows that 
% %
% % 
% $\lim_{|\alpha|\to\infty} {\rm spr}(p_{d, \alpha})=\infty.$
% %
% Let $\bar\alpha>0$ be such that ${\rm spr}(p_{d, \bar\alpha})>1.$
% %
% %
% Since $\alpha\mapsto {\rm spr}(p_{d, \alpha})$ is continuous, 
% % %
% % Furthermore, it follows from  the continuity of $\alpha\mapsto {\rm spr}(p_{d, \alpha})$ that there exist $\alpha_{d,0},\alpha_{d,1}>0$ such that
% % 
% % %
% it follows from Lemma \ref{propGIVT} with $f(\alpha) = {\rm spr}(p_{d, \alpha}),$ $a=0,$ $b=\bar\alpha,$ and $y=1$ that
% %
% there exist $\alpha_{d,0}, \alpha_{d,1}>0$ such that $0 < \alpha_{d,0} < \alpha_{d,1} < \bar\alpha,$ ${\rm spr}(p_{d, \alpha_{d,0}}) < 1 < {\rm spr}(p_{d, \alpha_{d,1}}),$ and, for all $\alpha \in(\alpha_{d,0},\alpha_{d,1}),$ ${\rm spr}(p_{d, \alpha_{d,0}}) < {\rm spr}(p_{d, \alpha}) < {\rm spr}(p_{d, \alpha_{d,1}}).$ 
% % 
% %
% Hence, {\it i}) holds.  A similar argument yields {\it ii}).
%
%
%
%
%
To prove  {\it iii}), note that, for all $\theta\in(0,\pi]$, $G(e^{\jmath \theta})\ne0,$ $W(e^{\jmath \theta})\ne0,$ and $G_d(e^{\jmath \theta})\ne0,$  and thus $L_d(e^{\jmath \theta})\ne0.$
Next, let $\theta \in (0, \pi],$ and note that
\begin{align*}
    \frac{\sin \theta}{\cos \theta - 1} 
    &= \frac{2 \sin \tfrac{\theta}{2} \cos \tfrac{\theta}{2}}{\cos^2 \tfrac{\theta}{2} - \sin^2 \tfrac{\theta}{2} - (\sin^2 \tfrac{\theta}{2} + \cos^2 \tfrac{\theta}{2})} \nn \\
    &= \frac{\cos \tfrac{\theta}{2}}{-\sin \tfrac{\theta}{2}} = \frac{\sin \tfrac{\pi+\theta}{2}}{\cos \tfrac{\pi+\theta}{2}},
\end{align*}
which implies 
\begin{equation*}
    \arg (\cos \theta - 1 + \jmath \sin \theta) =  \arg (\cos \tfrac{\pi+\theta}{2} + \jmath \sin \tfrac{\pi+\theta}{2}).
\end{equation*}
Hence
\begin{align}
    \angle W(e^{\jmath\theta})&=\arg (e^{\jmath\theta} - 1) - \arg (e^{\jmath\theta})\nn\\
    &= \arg (\cos \theta - 1 + \jmath \sin \theta)-\theta\nn\\
    % %
    &= \arg (\cos \tfrac{\pi+\theta}{2} + \jmath \sin \tfrac{\pi+\theta}{2})-\theta\nn\\
    &= \tfrac{\pi+\theta}{2}-\theta\nn\\
    &= \tfrac{\pi}{2} - \tfrac{\theta}{2}.\label{Wang} %\label{atanID}
\end{align}
Next, letting $d \in \BBN,$ it follows from \eqref{Wang} that
\begin{align}
    \angle L_d(e^{\jmath\theta}) &= \angle G(e^{\jmath\theta}) + \angle W(e^{\jmath\theta}) + \angle G_d(e^{\jmath\theta}) \nn \\
    &=  \angle G(e^{\jmath\theta}) + \tfrac{\pi}{2} - \tfrac{\theta}{2} - d\theta\nn \\
    &= \angle G(e^{\jmath\theta}) + \tfrac{\pi}{2}-(d+\half)\theta.\label{bubble}
\end{align}
% 

% \begin{align*}
%     \angle W(e^{\jmath\theta})&=\arg (e^{\jmath\theta} - 1) - \arg (e^{\jmath\theta})\nn\\
%     %
%     % &= \atann(\sin\theta,\cos\theta-1)-\theta\nn\\
%     % %
%     % &=? \frac{\pi+\theta}{2}-\theta\nn\\
%     %
%     %
%     %
%     %%%%%%%%%%%%%%%%%%%
%     &=\pi + {\rm atan} \left(\frac{\sin \theta}{\cos\theta - 1}\right) - \theta\nn\\
%     &=\pi+{\rm atan} \left({\rm cot} \left(-\tfrac{\theta}{2}\right)\right) - \theta \nn\\
%     &= \pi+{\rm atan} \left(-{\rm tan}\left(-\tfrac{\theta}{2} + \tfrac{\pi}{2}\right)\right) - \theta \nn\\
%     %%%%%%%%%%%%%%
%     %
%     %
%     %
%     %
%     &= \pi + \tfrac{\theta}{2} - \tfrac{\pi}{2} - \theta\nn\\ 
%     %
%     &= \tfrac{\pi}{2} - \tfrac{\theta}{2}. %\label{atanID}
% \end{align*}

%
% \begin{align*}
%     \angle W(e^{\jmath\theta})&={\rm atan} \left(\frac{  \sin \theta}{1- \cos \theta}\right)= {\rm atan} \left({\rm cot} \frac{\theta}{2}\right) \nn\\
%     &= {\rm atan} \left({\rm tan} \left(\frac{\pi}{2} - \frac{\theta}{2} \right)\right) = \frac{\pi}{2} - \frac{\theta}{2}. %\label{atanID}
% \end{align*}
%
%Gdf(e^jtheta) = 1-e^-jtheta = 1-cos+jsin
%angle Gdf(e^jtheta) =  atan(sin/(1-cos))
%tan theta/2 = (1-costheta)/sintheta 
%tan(pi/2-theta/2) = sin/(1-cos)
%angle of Gdf = atan(sintheta/(1-costheta)) = pi/2-theta/2
%
%
Now, let $\bar{d}\in\BBN$ satisfy
\begin{align}
    \max_{\theta\in[0,\pi] } \frac{\rmd}{\rmd\theta}\angle G(e^{\jmath\theta}) \le \bar{d}+\half.\nn
\end{align}
Therefore, for all $\theta\in(0,\pi]$ and $d>\bar{d},$
\begin{align}  
    \frac{\rmd}{\rmd\theta}\angle L_d(e^{\jmath\theta})
    &= \frac{\rmd}{\rmd\theta}\angle G(e^{\jmath\theta}) -d-\half\nn\\
    &\le  \max_{\theta\in(0,\pi] } \frac{\rmd}{\rmd\theta}\angle G(e^{\jmath\theta}) -d-\half\nn\\
    &\le \bar{d}+\half - d - \half
    <0.\nn
\end{align}

To prove {\it iv}), note that {\it iii}) implies that,
for all $d>\bar{d}$, $\angle L_d(e^{\jmath\theta})$ is a decreasing function of $\theta$ on $(0,\pi].$
Hence, for all $\alpha > 0,$  all crossings of the positive real axis by the Nyquist plot of $L_{d,\alpha}(e^{\jmath\theta}) = \alpha L_{d}(e^{\jmath\theta})$ as $\theta$ increases over the interval $(-\pi,\pi]$ occur from the first quadrant to the fourth quadrant.
%
%
%
%%%%%%%%%%%%%%%%%%%%%%%%%%%%%%%%%%%%%
%
%
Next, note that, for all $d> \bar{d}$ and $\theta\in(0,\pi],$ $|L_{d,\alpha}(e^{\jmath \theta})| = \alpha |L_d(e^{\jmath \theta})|$ is a increasing function of $\alpha$ on $(0,\infty),$   
and that, for all $d>\bar{d}$ and $\alpha>0,$ since all of the poles of $L_{d, \alpha}$ are in the open unit disk, it follows that ${\rm spr}(p_{d,\alpha}) > 1$ if and only if the number of clockwise encirclements of $1+0\jmath$ of the Nyquist plot of $L_{d,\alpha}(e^{\jmath \theta})$ over $\theta\in(-\pi, \pi]$ is at least one.
Therefore, for all $d > \bar{d}$ and $\alpha_0, \alpha_1 >0$ such that ${\rm spr}(p_{d,\alpha_0}) > 1$
and $\alpha_1 > \alpha_0,$ the Nyquist plot of $L_{d,\alpha_1}(e^{\jmath \theta})$ over $\theta\in(-\pi,\pi]$ has at least one clockwise encirclement of $1+0\jmath$.  %its not asymptotically stable
Furthermore, for all $d > \bar{d}$ and $\alpha_0, \alpha_1 >0$ such that ${\rm spr}(p_{d,\alpha_0}) < 1$
and  $\alpha_1 < \alpha_0,$ the Nyquist plot of $L_{d,\alpha_1}(e^{\jmath \theta})$ over $\theta\in(-\pi,\pi]$ has zero encirclements of $1+0\jmath.$
Hence, {\it i}) implies that there exists a unique $\alpha_{d,\rmu}>0$ such that ${\rm spr}(p_{d,\alpha_{d,\rmu}}) = 1$, for all $\alpha \in [0,\alpha_{d,\rmu}),$ $\spr(p_{d,\alpha})<1$, and, for all $\alpha \in [\alpha_{d,\rmu},\infty),$ $\spr(p_{d,\alpha})\ge1.$
Similarly, {\it ii}) implies that there exists a unique $\alpha_{d,\rml}<0$ such that ${\rm spr}(p_{d,\alpha_{d,\rml}}) = 1$, for all $\alpha \in (\alpha_{d,\rml},0],$ $\spr(p_{d,\alpha})<1$ and, for all $\alpha \in (-\infty,\alpha_{d,\rml}),$ $\spr(p_{d,\alpha})>1.$
Hence, {\it iv}) holds. 
To prove  {\it v}),
let $\alpha\in\BBR$ and $d\ge\bar{d}.$
%
%and define
%
%$\roots(p_{d,\alpha})\isdef\{\lambda\colon p_{d,\alpha}(\lambda)=0\}.$
%
%$\roots(p_{d,\alpha})\isdef\{\lambda\colon \lambda \in \roots(p_{d,\alpha})\}.$
%
%$\roots(p_{d,\alpha}) \isdef \roots(p_{d,\alpha}).$
%
%\roots(p_{d,\alpha}).$
%
Note that $\roots(p_{d,\alpha})$ has at most $n+d+1$ elements and that $\lambda\in\roots(p_{d,\alpha})$ if and only if 
$L_{d,\alpha}(\lambda)=1.$
Now, let $\lambda =\rho  e^{\jmath \theta}\in\roots(p_{d,\alpha}),$ where $\rho\in[0,\infty)$ and $\theta\in(-\pi,\pi].$ 
%
%$\lambda =\rho  e^{\jmath \theta}.$
%
%
%
Writing $G(z) =  \frac{K\prod_{k = 1}^{m} (z - z_k)}{\prod_{k = 1}^{n} (z - p_k)}$, it follows from $L_{d,\alpha}(\lambda)=1$ that
\begin{align}
    |\alpha| &= \frac{|\lambda^{d+1}|\prod_{k = 1}^{n} |\lambda - p_k|}{|K||\lambda - 1| \prod_{k = 1}^{m} |\lambda - z_k|} \nn \\
    &= \rho^{d+n-m}  \frac{\prod_{k = 1}^{n} |e^{ \jmath \theta} - \frac{p_k}{ \rho}|}{|K||e^{ \jmath \theta} - \frac{1}{\rho}| \prod_{k = 1}^{m} |e^{ \jmath \theta} - \frac{z_k}{ \rho}|}.
     \label{eqRhoD}
\end{align}

The case where $\alpha>0$ is considered henceforth.
Let $\SA_{d,+}$ denote the set of all $\alpha > 0$ such that $\roots(p_{d,\alpha})$ has at least one element with magnitude 1.
%
%
%
%new statement i
It follows from {\it i}) that $\SA_{d,+}$ is not empty.
Furthermore,  {\it iv}) implies that $\SA_{d,+}\subseteq [\alpha_{d, \rmu},\infty).$
Finally, {\it i}) and {\it iv}) imply that ${\rm spr}(p_{d,\alpha_{d,\rmu}}) = 1,$ and thus $\alpha_{d, \rmu} = \min \SA_{d,+}.$

For all $\alpha\in\SA_{d,+},$  let $\theta_\alpha\in(0,\pi] $ satisfy $p_{d,\alpha}(e^{\jmath\theta_\alpha})=0.$  
It thus follows from \eqref{eqRhoD} with $\rho=1$ that, for all $\alpha\in\SA_{d,+},$
\begin{align}
    \alpha = g(\theta_\alpha),\label{alpha0Eq}
\end{align}    
where $g\colon(0,\pi]\to(0,\infty)$ is defined by
\begin{equation*} 
    g(\theta)\isdef \frac{\prod_{k = 1}^{n} |e^{ \jmath \theta} - p_k|}{|K||e^{ \jmath \theta} - 1| \prod_{k = 1}^{m} |e^{ \jmath \theta} - z_k|}.
    \label{alphaRho1}
\end{equation*}
Since $g$ is continuous and $\lim_{\theta\downarrow0}g(\theta) =\infty,$ it follows that $g$ has a global minimizer.  %$\theta_\infty\in(0,\pi].$
Hence, define the set of minimizers of $g$ by
\begin{equation}
    M\isdef \{\theta\in(0,\pi]\colon g(\theta) =  \alpha_\infty\},\nn
\end{equation}
where 
%
%
%
%%%%%%%%%%%%%%%%%%%%%%%%%%%%%%%%%
% Let $\Theta_0$ be any closed interval in $[0, \pi]$ in which $f_\alpha \colon [0, \pi] \to [0, \infty)$ is continuous.
% %
% %
% %
% %
% %
% %
% %
% %JJJJ: This is in case a zero is located at the unit circle, which would make the function continuous by intervals. I still need to think of a more precise way of explaining this.
% %
% For all $\Theta_0,$ it follows that $f_\alpha^{-1} (\{[0, \infty)\})$ is compact.
% %
% Since $\cup\Theta_0 = [0, \pi],$ then it follows that the minimum of $f_\alpha \colon [0, \pi] \to [0, \infty)$ exists.
%%%%%%%%%%%%%%%%%%%%%%%%%%%%%
%
%
\begin{align*}
    \alpha_\infty &\isdef  \min_{\theta \in (0, \pi]}g(\theta)\nn \\
    &= \min_{\theta \in (0, \pi]} \frac{\prod_{k = 1}^{n} |e^{ \jmath \theta} - p_k|}{|K||e^{ \jmath \theta} - 1| \prod_{k = 1}^{m} |e^{ \jmath \theta} - z_k|} \nn \\
    &= \min_{\theta\in(0,\pi]}\left|\frac{D(e^{\jmath\theta})}{(e^{\jmath\theta}-1)N(e^{\jmath\theta})}\right|. \label{AMagEq}
\end{align*}
Hence, the minimum in \eqref{alphaInfEq} exists, is positive, and is independent of $d.$ 
Furthermore, for all $\alpha\in\SA_{d,+},$ $\alpha_\infty \le g(\theta_\alpha) = \alpha,$ and thus 
$\alpha_\infty \le \min \SA_{d,+}.$

Next, we show that there exists $r_0\in \BBZ$ such that $\angle L_d(1) + 2r_0\pi \in (0, \pi)$ and that, for all $d\geq\bar{d},$ there exists $r_d \in \BBZ$ such that $\angle L_d(e^{\jmath \pi}) + 2r_d\pi \in [-\pi, 0]$.
We now consider the case where $G(1) > 0$;  the case $G(1) < 0$ is addressed by the case where $\alpha < 0.$
%
% Then, for all $\theta \in (0, \pi],$
% %
% there exists $r\in\BBZ$ such that $\angle L_{d}(e^{\jmath \theta}) + 2 r\pi \in [-\pi, \pi).$
%
%
Let $r_0\in\BBZ$ satisfy $\angle G(1) + 2  r_0\pi = 0,$ and
note that  \eqref{bubble} implies that, for all $d\ge \bar{d},$  $\angle L_d(1)\isdef\lim_{\theta\downarrow0}\angle L_d(e^{\jmath\theta}) = \half\pi+\angle G(1).$  It thus follows that 
\begin{equation}
    \angle L_d(1) + 2 r_0\pi  = \half \pi\in(0,\pi). \label{Lej0Eqn}
\end{equation}
%
%
%
%
% It thus follows from \eqref{Lej0Eqn} that there exists $r_0\in\BBZ$ such that
% %
% \begin{equation}
%     \angle L_d(1) + 2r_0\pi \in (0, \pi). \label{Lej0Eqn1}
% \end{equation}
% %

Next,  \eqref{bubble} with $\theta=\pi$ implies that, for all $r\in\BBZ,$  
\begin{equation}
    \angle L_d(-1) + 2r \pi = \angle G(-1) - d \pi+ 2r \pi. \label{LejPiEqn}
\end{equation}
Let $r_\rmm\in\BBZ$ satisfy $\angle G(-1) + 2 r_\rmm\pi \in \{-\pi, 0\}.$
Consider the case where $G(-1) > 0,$ 
and thus $\angle G(-1) + 2 r_\rmm\pi = 0.$
Define $r_{d,\rmp} \isdef r_\rmm+\lfloor\frac{d}{2}\rfloor$.
In the case where $d$ is even, it  follows from  \eqref{LejPiEqn} with $r = r_{d,\rmp}$ that $\angle L_d(-1) +  2r_{d,\rmp}\pi = 0$.
Likewise, in the case where $d$ is odd, $\angle L_d(-1) + 2r_{d,\rmp}\pi = -\pi$.
%
%
%
%
%
%
%It follows from \eqref{LejPiEqn} that, assuming $G(e^{\jmath \pi}) = G(-1) > 0$ and, thus, $\angle G(-1) = 0,$ and replacing $r$ with $\lfloor\half d \rfloor + r_\rmm,$ in the case where $d$ is even, $\angle L_d(-1) + 2r_d\pi = 0$, and, in the case where $d$ is odd, $\angle L_d(-1) + 2r_d\pi = -\pi.$
%
Hence, 
\begin{equation}\label{rdpEq}
\angle L_d(-1) + 2r_{d,\rmp} \pi \in \{-\pi, 0\}.
\end{equation}
%
%
%Next, in the case where $G(-1) < 0$ and thus $\angle G(-1) + 2\pi r_\rmm = -\pi,$ it follows from \eqref{LejPiEqn} with $r=\lfloor\half (d+1) \rfloor + r_\rmm$ that, in the case where $d$ is even, $\angle L_d(-1) + 2r_d\pi = -\pi$, and, in the case where $d$ is odd, $\angle L_d(-1) + 2r_d\pi = 0.$
%
%it follows from \eqref{LejPiEqn} that, assuming $G(-1) < 0$ and, thus, $\angle G(-1) + 2\pi r_\rmm = -\pi,$ and replacing $r$ with $\lfloor\half (d+1) \rfloor + r_\rmm,$ in the case where $d$ is even, $\angle L_d(-1) + 2r_d\pi = -\pi$, and, in the case where $d$ is odd, $\angle L_d(-1) + 2r_d\pi = 0.$
%
%
Similarly, in the case where $G(-1) < 0,$ define $r_{d,\rmn} \isdef  r_\rmm+\lfloor \tfrac{d+1}{2} \rfloor$ so that
\begin{equation}\label{rdnEq}
\angle L_d(-1) + 2r_{d,\rmn} \pi \in \{-\pi, 0\}.
\end{equation}
Note that, in the case where $d$ is even, $r_{d,\rmp} = r_{d,\rmn} = r_\rmm+\tfrac{d}{2},$
whereas, in the case where $d$ is odd, $r_\rmm+\tfrac{d-1}{2} = r_{d,\rmp} < r_{d,\rmn} = r_\rmm+\tfrac{d+1}{2}.$
%
%Hence, for all integers $r \in [r_{d,\rmp},r_{d,\rmn}],$
%
%what would happen if we change line 3057 to be:
%
%Hence, for all $r \in {r_{d,\rmp},r_{d,\rmn}},$
%
%it would be true, but would it suffice to reach 25 and 26?  why do you need to mention the entire INTERVAL----(which might be just one point anyway)?
%
%I will need to leave, so pls think about this.  I will check when I get back thanks
%
%JJJ: I think it would be ok. 25 and 26 would remain true with this change. Making this point for the entire interval is not necessary. I will be away for a while as well. Please, make any necessary changes and leave comments where needed. Thank you.
%
%
% \begin{equation}
%     2r_\rmm + d - 1  \leq 2 r \leq 2r_\rmm + d + 1. \label{RdBound0}
%     %2 r_{d,\rmo} \in [2r_\rmm + d - 1, \ 2r_\rmm + d + 1]. \label{RdBound0}
% \end{equation}
%
%
%I will check this later today, thanks
%I need to read it carefully to be sure I understand the logic.  thanks
%
%
It thus follows from \eqref{rdpEq} and \eqref{rdnEq}
that, in both cases, that is, $G(-1)>0$ and $G(-1)<0,$ there exists $r_d\in\{r_{d,\rmp}, r_{d,\rmn}\}$ such that 
\begin{equation}
    2r_\rmm + d - 1  \leq 2 r_d \leq 2r_\rmm + d + 1 \label{RdBound}
    %2 r_d \in [2r_\rmm + d - 1, \ 2r_\rmm + d + 1]. \label{RdBound}
\end{equation}
and
%
%JJJ: I added the subset since that is the conclusion we said we were going to arrive to in line 2964
%
\begin{equation}\label{rdEq}
\angle L_d(-1) + 2r_{d} \pi \in \{-\pi, 0\}\subset [-\pi, 0].
\end{equation}
%

% %
% Thus, it follows from \eqref{rdpEq} and \eqref{rdnEq} that there exists $r_d \in \{r_{d,\rmn}, r_{d,\rmp}\}$ such that
% %
% \begin{equation}
%     d-1 + 2r_\rmm \leq 2 r_d \leq d+1 + 2r_\rmm, \label{RdBound}
% \end{equation}
% %
% and
% %
% \begin{equation}
%     \angle L_d(e^{\jmath \pi}) + 2r_d \pi \in [-\pi,0]. \label{rdEq}
% \end{equation}
% %

% In the case where $\angle G(-1) = 0,$ $r = \lfloor\half d \rfloor$ yields $\angle L_{d,\alpha}(-1) + 2r\pi = 0$ for even $d$ and $\angle L_{d,\alpha}(-1) + 2r\pi = -\pi$ for odd $d.$ 
% %
% Hence, for $\angle G(-1) = 0,$ $r_d = \lfloor\half d \rfloor.$
% %
% In the case where $\angle G(-1) = -\pi,$ $r = \lfloor\half (d+1) \rfloor$ yields $\angle L_{d,\alpha}(-1) + 2r\pi = -\pi$ for even $d$ and $\angle L_{d,\alpha}(-1) + 2r\pi = 0$ for odd $d.$
% %
% Hence, for $\angle G(-1) = -\pi,$ $r_d = \lfloor\half (d+1) \rfloor.$
% %

%\clearpage 
%Let $d > \bar{d}.$  
%
%
%, for all $\theta \in (0, \pi],$ $|L_d(e^{\jmath \theta})| \neq 0.$
%
%Hence,
%
Next, let $d \geq \bar{d}.$ Then, \textit{iii}) implies that,  $\theta\mapsto\angle L_d(e^{\jmath \theta})$ is continuous and decreasing on $(0, \pi].$
It thus follows from \eqref{Lej0Eqn} that
\begin{equation}
    r_0 = \frac{1}{4}-\frac{1}{2\pi}\angle L_d(1)\label{r0is14}
\end{equation}
and from \eqref{rdEq} that
$
    r_d \in\{r_{d,{\rm min}},r_{d,{\rm max}}\},
$
where 
%
%
%ok?---JJJ: Yes, this is ok.
%
%
\begin{align*}
    r_{d,{\rm min}} & \isdef -\frac{1}{2}-\frac{1}{2\pi}\angle L_d(-1),\\
    r_{d,{\rm max}} & \isdef -\frac{1}{2\pi}\angle L_d(-1).
\end{align*}
Since $\angle L_d(-1) = \angle L_d(e^{\jmath \pi}) < \angle L_d(1)$, it follows from \eqref{r0is14} that
\begin{align*}
r_0 
% &= \frac{1}{4}-\frac{1}{2\pi}\angle L_d(1)\nn\\
%
&< \frac{1}{4}-\frac{1}{2\pi}\angle L_d(-1) = \frac{1}{4} + r_{d,{\rm max}} = \frac{3}{4} + r_{d,{\rm min}}\\
&< 1 + r_{d,{\rm min}} < 1 + r_{d,{\rm max}},
\end{align*}
and thus $r_0 < r_d+1,$ which implies that $r_0\le r_d.$

Then, let $d \geq \bar{d}.$ Since \eqref{Lej0Eqn} implies that $\angle L_d(1) + 2 r_0\pi > 0$ and \eqref{rdEq} implies that $\angle L_d(-1) + 2r_{d} \pi \leq 0,$ it follows that, for all $r\in \{r_0, \ldots, r_d\},$
\begin{equation}\label{Ld1_r_EQ}
    \angle L_d(1) + 2r\pi > 0,
\end{equation}
and
\begin{equation}\label{Ld_1_r_EQ}
    \angle L_d(-1) + 2r\pi \leq 0.
\end{equation}
Thus, since $\theta \to \angle L_d(e^{\jmath \theta})$ is decreasing and continuous on $(0,\pi],$ it follows from \eqref{Ld1_r_EQ},  \eqref{Ld_1_r_EQ}, and the intermediate value theorem that, for all $r\in \{r_0, \ldots, r_d\},$ there exists a unique $\theta_{r,d} \in (0, \pi]$ such that
\begin{equation}
    \angle L_d (e^{\jmath \theta_{r,d}}) + 2r \pi = 0. \label{angLREq}
\end{equation}
Furthermore, let $r_\rml,r_\rmh\in\{r_0, \ldots, r_d\}$ such that $r_\rml\leq r_\rmh,$
and let $\theta_{r_\rml,d},\theta_{r_\rmh,d}\in(0,\pi]$ satisfy
\begin{equation} \label{angLREqEx}
    \angle L_d (e^{\jmath \theta_{r_\rml,d}}) + 2r_\rml \pi = \angle L_d (e^{\jmath \theta_{r_\rmh,d}}) + 2r_\rmh \pi = 0.
\end{equation}
In the case where $r_\rml = r_\rmh,$ \eqref{angLREqEx} implies that $\theta_{r_\rml,d} = \theta_{r_\rmh,d}$. In the case where $r_\rml<r_\rmh,$ \eqref{angLREqEx} implies that $\angle L_d (e^{\jmath \theta_{r_\rmh,d}}) < \angle L_d (e^{\jmath \theta_{r_\rml,d}}),$ and, since $\theta \to \angle L_d(e^{\jmath \theta})$ is decreasing on $(0,\pi],$ $\theta_{r_\rml,d} < \theta_{r_\rmh,d}.$
Hence, in the case where $r_0 = r_d,$ it follows that $\theta_{r_0,d} = \theta_{r_d,d},$ and, in the case where $r_0 < r_d,$ it follows that $\theta_{r_0,d} < \theta_{r_0+1,d} < \cdots < \theta_{r_d,d}.$
%

%
%Next, for all $d\geq \bar{d},$ define
%
% \begin{align}
%     \Theta_{d,+} \isdef \{\theta \in (0,\pi] \colon &\text{there exists }\alpha \in \SA_{d,+} \nn \\ 
%     &\text{ such that } L_{d,\alpha}(e^{\jmath \theta}) = 1\}. \nn
% \end{align}
%
%Note that, for all $d\geq \bar{d},$ $\theta \in \Theta_{d,+}$ if and only if there exists $\alpha \in \SA_{d,+}$ such that $L_{d,\alpha}(e^{\jmath \theta}) = 1.$
%
Next, let $d\geq \bar{d},$ and let $r \in \{r_0, \ldots, r_d\}$  and $\theta_{r,d} \in (0, \pi]$ satisfy \eqref{angLREq},
so that
$\angle L_d (e^{\jmath \theta_{r,d}})$ is an integer multiple of $2\pi.$ Therefore, 
$L_d (e^{\jmath \theta_{r,d}})$ is a positive number, and thus 
$
    L_{d,\alpha_r} (e^{\jmath \theta_{r,d}}) = 1,
$
where $\alpha_r \isdef 1/L_d (e^{\jmath \theta_{r,d}})>0.$
Therefore, $p_{d,\alpha_r}(e^{\jmath \theta_{r,d}})=0,$ and thus 
$\alpha_r \in \SA_{d,+}$,
which implies that
\begin{equation*}
    \theta_{r,d} \in \Theta_{d,+}, \label{thetaInThetadEq}
\end{equation*}
where
\begin{align}
    \Theta_{d,+} \isdef \{\theta \in (0,\pi] \colon &\text{there exists }\alpha \in \SA_{d,+} \nn \\ 
    &\text{ such that } L_{d,\alpha}(e^{\jmath \theta}) = 1\}. \nn
\end{align}
Now suppose that, for all $d \geq \bar{d},$ there exists $\theta \in \Theta_{d,+}\backslash \{\theta_{r_0,d}, \theta_{r_0+1,d}, \ldots, \theta_{r_d,d}\}.$
Hence there exists $r \in \BBZ\backslash\{r_0, \ldots, r_d\}$ such that $\angle L_d(e^{\jmath \theta}) + 2r\pi = 0.$
In the case where $r < r_0,$ it follows from \eqref{Lej0Eqn} that
\begin{equation}\label{Ld1_Eq_Suff}
    \angle L_d(1) + 2r\pi \leq \angle L_d(1) + 2(r_0-1)\pi = -\tfrac{3}{2} \pi.
\end{equation}
Since $\theta \to \angle L_d(e^{\jmath \theta})$ is decreasing on $(0,\pi],$ \eqref{Ld1_Eq_Suff} implies that, for all $\theta\in(0,\pi],$
\begin{equation}\label{Ld_1_Eq_Suff}
    \angle L_d(e^{\jmath \theta}) + 2r\pi < -\tfrac{3}{2}\pi< 0 = L_d(e^{\jmath \theta}) + 2r\pi,
\end{equation}
which is a contradiction.
Hence, $r> r_d.$
Similarly, supposing that $r > r_d$ also leads to a contradiction.
%
% for all $\theta\in(0,\pi],$ \eqref{Ld_1_Eq_Suff} follows from \eqref{rdEq},
%
%that
% %
% \begin{align}\label{Ld_1_Eq_Suff}
%     \angle L_d(-1) + 2r\pi &\geq \angle L_d(-1) + 2(r_d+1)\pi \in \{\pi, 2\pi\} \nn \\
%     &\geq \pi.
% \end{align}
% %
% Since $\theta \to \angle L_d(e^{\jmath \theta})$ is decreasing on $(0,\pi],$ \eqref{Ld_1_Eq_Suff} implies that,
% %
% for all $\theta\in(0.\pi],$
% %
% \begin{equation*}
%     \angle L_d(e^{\jmath \theta}) + 2r\pi \neq 0,
% \end{equation*}
% %
% which is a contradiction.
%
Therefore, for all $d\ge \bar{d},$  $\Theta_{d,+}= \{\theta_{r_0,d}, \theta_{r_0+1,d}, \ldots, \theta_{r_d,d}\}.$

%\clearpage

Next, for all $d \geq \bar{d}$ and $r \in \{r_0, \ldots, r_d\}$, adding $2\pi r$ to both sides of \eqref{bubble} with $\theta=\theta_{r,d},$ it follows from \eqref{angLREq} that
%
%For all integers $r \in [r_0, r_d],$ there exists a unique $\theta_r \in (0, \pi]$ such that \eqref{angLREq} holds and, thus, it follows from \eqref{anglaL_alphaEqn} that, by replacing $\theta$ with $\theta_r,$
%
\begin{align}
    \theta_{r,d} = \frac{4r\pi+2\angle G(e^{ \jmath \theta_{r,d}}) + \pi}{2d+1}. \label{theta_du_Eq1}
\end{align}
Then, \eqref{theta_du_Eq1} implies that, for all $d\geq\bar{d}$ and $r \in \{r_0+1, \ldots, r_d\},$
\begin{equation}\label{deltaThetaEqRLim0}
    \theta_{r,d} - \theta_{r-1,d} = \frac{2}{2d + 1} [2\pi + \angle G(e^{\jmath \theta_{r,d}}) - \angle G(e^{\jmath \theta_{r-1,d}})].
\end{equation}
% %
% and thus
% \begin{align}
%   \lim_{d \to \infty} &(\theta_{r,d} - \theta_{r-1,d}) \nn \\
%   &= \lim_{d \to \infty} \left(\frac{2}{2d + 1} [2\pi + \angle G(e^{\jmath \theta_{r,d}}) - \angle G(e^{\jmath \theta_{r-1,d}})]\right) \nn \\
%   &= 0.\label{deltaThetaEqRLim}
% \end{align}
%

%Let $d \geq \bar{d}.$
%
In the case where $r = r_0,$ it follows from \eqref{theta_du_Eq1} with $r = r_0$ that, for all $d\geq \bar{d},$
\begin{align}
    \theta_{r_0,d} = \frac{4r_0\pi+2\angle G(e^{ \jmath \theta_{r_0,d}}) + \pi}{2d+1}. \label{theta0_du_Eq}
\end{align}
It thus follows from \eqref{theta0_du_Eq} that, for all $\theta \in (0, \pi],$ there exists $d_{\theta,\rml} \geq \bar{d}$ such that, for all $d\geq d_{\theta,\rml},$
\begin{equation}
    \theta_{r_0,d} < \theta. \label{theta0LowBound}
\end{equation}
In the case where $r = r_d,$ it follows from \eqref{RdBound} and \eqref{theta_du_Eq1} with $r = r_d$ that, for all $d\geq \bar{d},$
\begin{align}
    \theta_{r_d,d} &=\frac{4r_d \pi + 2\angle G(e^{ \jmath \theta_{r_d,d}}) + \pi}{2d+1} \nn \\
    &\geq \frac{2(2r_\rmm + d - 1) \pi + 2\angle G(e^{ \jmath \theta_{r_d,d}}) + \pi}{2d+1} \nn \\
    &= \pi - \frac{2\pi - 2(\angle G(e^{ \jmath \theta_{r_d,d}})+2\pi r_\rmm)}{2d+1}.\label{thetaRd_du_Eq}
\end{align}
% %
% Since $\theta_{r_d,d}\leq\pi,$ it follows from \eqref{thetaRd_du_Eq} that $\angle G(e^{ \jmath \theta_{r_d,d}})+2\pi r_\rmm \leq \pi.$
%
Hence, for all $\theta \in (0, \pi),$
\begin{align}
  -\frac{2\pi - 2(\angle G(e^{ \jmath \theta_{r_d,d}})+2\pi r_\rmm)}{2d+1}+\pi-\theta \le \theta_{r_d,d}-\theta,\nn 
\end{align}
which implies that there exists $d_{\theta,\rmr} \geq \bar{d}$ such that, for all $d\geq d_{\theta,\rmr},$
\begin{equation}
    \theta_{r_d,d} > \theta. \label{thetaRdUppBound}
\end{equation}
Furthermore,  \eqref{thetaRd_du_Eq} implies that
\begin{equation}
    \pi\ge\lim_{d \to \infty} \theta_{r_d,d} \ge \pi - \lim_{d \to \infty} \tfrac{2\pi - 2(\angle G(e^{ \jmath \theta_{r_d,d}})+2\pi r_\rmm)}{2d+1} = \pi. \nn
\end{equation}
Hence, 
\begin{equation}\label{thetaRdInfty}
    \lim_{d \to \infty} \theta_{r_d,d} = \pi.
\end{equation}

Next, let $\theta_\infty \in M.$
We first consider the case where $\theta_\infty \in (0,\pi).$
It follows from \eqref{theta0LowBound} and \eqref{thetaRdUppBound} with $\theta = \theta_{\infty}$ that, for all $d\geq\max\{d_{\theta_\infty,\rml}, d_{\theta_\infty,\rmr}\},$ there exists $r \in \{r_0+1, \ldots, r_d \}$ such that
\begin{equation}\label{thetaRDiffBound1}
    \theta_{r-1,d} \leq \theta_\infty \leq \theta_{r,d}.
\end{equation}
It follows from \eqref{deltaThetaEqRLim0} and \eqref{thetaRDiffBound1} that, for all $\varepsilon > 0,$ there exists $d_{\theta_\infty,\rmm}\geq  \max\{d_{\theta_\infty,\rml}, d_{\theta_\infty,\rmr}\}$ such that, for all $d\geq d_{\theta_\infty,\rmm},$ there exists $r \in \{r_0+1, \ldots, r_d \}$ such that
\begin{equation}\label{thetaRDiffBound2}
     0 \leq \theta_\infty - \theta_{r-1,d} \leq \theta_{r,d} - \theta_{r-1,d} < \varepsilon
\end{equation}
and
\begin{equation}\label{thetaRDiffBound3}
     0 \leq \theta_{r,d} - \theta_\infty \leq \theta_{r,d} - \theta_{r-1,d} < \varepsilon.
\end{equation}
%
%Now define the sequence ${(\psi_{\theta_\infty,\tilde{d}})}_{\tilde{d} = 0}^{\infty}\subset(0,\pi]$ where, for all $\tilde{d}\ge0,$  
%
Now, for all $\tilde{d}\ge0,$ define 
\begin{equation*}
    \psi_{\theta_\infty,\tilde{d}} \isdef \underset{{\theta} \in \Theta_{\tilde{d},+}}{\rm argmin} |\theta_\infty - {\theta}| \in (0,\pi].
\end{equation*}
%
%
%Hi juan, are you there?  in the defionition of psi, you use the index k.  However, k is like the time step.  Adctually this index is reallly d, but I realize you cannot use d.  Can you try overline{d}?  thanks--Will do. Just give me a moment.We already use $\bar{d}$ in this document. How about \tilde{d} ?
%
%
%
It follows from \eqref{thetaRDiffBound1} that, for all $\tilde{d} \geq \max\{d_{\theta_\infty,\rml}, d_{\theta_\infty,\rmr}\},$ 
%
%should theta be theta_infty???---Yes, I already corrected it.
%
%
%
there exists $r \in \{r_0+1, \ldots, r_{\tilde{d}} \}$ such that
\begin{equation*}
    \psi_{\theta_\infty,\tilde{d}} \in \{\theta_{r-1,\tilde{d}}, \theta_{r,\tilde{d}}\}, \label{psiThetaK}
\end{equation*}
%
%Furthermore, it follows from \eqref{thetaRDiffBound2}, \eqref{thetaRDiffBound3}, and \eqref{psiThetaK} that, for all $\varepsilon > 0,$ there exists $d_{\theta_\infty,\rmm}\geq  \max\{d_{\theta_\infty,\rml}, d_{\theta_\infty,\rmr}\}$ such that, for all $\tilde{d}\geq d_{\theta_\infty,\rmm},$ there exists $r \in \{r_0+1, \ldots, r_{\tilde{d}} \}$ such that
%
and thus, for all $\varepsilon > 0,$ \eqref{thetaRDiffBound2} and \eqref{thetaRDiffBound3} imply 
\begin{equation*}
    |\psi_{\theta_\infty,\tilde{d}} - \theta_\infty| \in \{\theta_\infty - \theta_{r-1,\tilde{d}}, \theta_{r,\tilde{d}} - \theta_\infty\} < \varepsilon.
\end{equation*}
Hence,
\begin{equation}\label{thetaLimDiffLast}
    \lim_{\tilde{d} \to \infty} \psi_{\theta_\infty,\tilde{d}} = \theta_\infty\in(0,\pi).
\end{equation}
In the case where $\theta_\infty = \pi,$ \eqref{thetaRdInfty} implies  
\begin{equation}\label{thetaLimDiffpi}
    \lim_{\tilde{d} \to \infty} \psi_{\pi,\tilde{d}} = \pi.
\end{equation}
Hence, \eqref{thetaLimDiffLast} and \eqref{thetaLimDiffpi} imply 
%
%should we say that it follows from (42) AND the limit in the case of theta_infty = pi???  Dont you need BOTH limits to conlude 43?
% JJJ: Yes, we need both. 
%
%
%
\begin{equation}\label{thetaBarLim}
    \lim_{\tilde{d} \to \infty} \psi_{\theta_\infty,\tilde{d}} = \theta_\infty\in(0,\pi].
\end{equation}
Since \eqref{alpha0Eq} implies that, for all $d \geq \bar{d},$ $\alpha_{d,\rmu} =\underset{\theta \in \Theta_{d,+}}{\min}  g(\theta),$ and, for all $\theta_\infty \in M,$ $\alpha_\infty = g(\theta_\infty),$ it follows from \eqref{thetaBarLim} that
\begin{equation}
    \lim_{d\to\infty} \alpha_{d,\rmu} = \alpha_\infty.  \label{alphaduLim}
\end{equation}
Similarly, in the case where $\alpha < 0,$  
\begin{equation}
    \lim_{d\to\infty} -\alpha_{d,\rml} = \alpha_\infty. \label{alphadlLim}
\end{equation}
Finally, \eqref{alphaduLim} and \eqref{alphadlLim} imply  
\begin{equation}
     \lim_{d\to\infty} -\alpha_{d,\rml} = \lim_{d\to\infty} \alpha_{d,\rmu} = \min_{\theta \in (0, \pi]}\left|\frac{D(e^{\jmath\theta})}{(e^{\jmath\theta}-1)N(e^{\jmath\theta})}\right|.\tag*{\mbox{$\square$}}
\end{equation}

\begin{prop}
Let $\alpha\in\BBR,$ $d\geq 0,$ and $\theta\in(0,\pi],$ and assume that
$p_{d,\alpha}(e^{\jmath\theta})=0$.
Then,
\begin{equation}\label{propAlphaTheta}
    \alpha = \frac{e^{\jmath(d+1) \theta}}{(e^{\jmath\theta}-1)G(e^{\jmath\theta})}.
\end{equation}
Furthermore, writing $G^{-1}(e^{\jmath\theta}) =  a + b\jmath,$ where $a,b\in\BBR,$ it follows that
\begin{equation}\label{propbthetaa}
    b = - a \frac{\sin d\theta - \sin\, (d+1)\theta}{\cos d\theta - \cos\, (d+1)\theta}
\end{equation}
and
\begin{equation}\label{propAlphaTheta2}
     \alpha %= \frac{ f}{2 - 2 \cos \theta} 
    = \frac{ a}{\cos d\theta - \cos\, (d+1)\theta}.
\end{equation}
\end{prop}

\smallskip

{\bf Proof.} 
 \eqref{propAlphaTheta} follows from $p_{d,\alpha}(e^{\jmath\theta}) = 0.$
Furthermore, \eqref{propAlphaTheta} implies that
\footnotesize 
\begin{align}
    {\alpha = \frac{[(\cos d\theta - \cos\, (d+1)\theta) + \jmath(\sin d\theta - \sin\, (d+1)\theta)] G^{-1}(e^{\jmath\theta})}{2 - 2\cos(\theta)},}
\end{align}
\normalsize
and thus
\begin{align} \label{alphaThetafg}
    \alpha = \frac{ f + \jmath  g}{2 - 2 \cos \theta},
\end{align}    
where
\small
\begin{align}   
    {f \isdef  a[\cos d\theta - \cos\, (d+1) \theta] -  b[\sin d\theta - \sin\, (d+1)\theta],}\label{fTheta}\\
    {g \isdef   b[\cos d\theta - \cos\, (d+1) \theta] +  a[\sin d\theta - \sin\, (d+1)\theta].}\label{gTheta}
\end{align}
\normalsize
Since $\alpha$ is real, \eqref{alphaThetafg} implies that $g=0,$ and thus  
%, for all $\theta\in(0,\pi]$ such that $e^{\jmath \theta}$ is a root of $p_{d,\alpha},$
% \begin{equation}
%     g = 0. \label{gTheta0}
% \end{equation}
%
\eqref{gTheta} implies \eqref{propbthetaa}.
Next, combining \eqref{propbthetaa} with \eqref{fTheta} yields
\small
\begin{align}\label{fTheta0}
%\begin{split}
     f & =  a \frac{\left[\cos d\theta - \cos\, (d+1)\theta\right]^2 + \left[\sin d\theta - \sin\, (d+1)\theta\right]^2}{\cos d\theta - \cos\, (d+1)\theta} \nn\\
    & =  a \frac{2 - 2 \cos d\theta \cos\, (d+1)\theta - 2 \sin d\theta \sin\, (d+1)\theta}{\cos d\theta - \cos\, (d+1)\theta} \nn\\
    & =  a \frac{2 - 2 \cos \theta}{\cos d\theta - \cos\, (d+1)\theta}.
%\end{split}
\end{align}
\normalsize
Finally, combining $g=0$ and \eqref{fTheta0} with \eqref{alphaThetafg} yields \eqref{propAlphaTheta2}.  \hfill$\square$

\begin{exam}
\label{ex_2_2}
Let $G(z) = \frac{1}{z + p}$, where $p \in (-1, 1)$,  and let $e^{\jmath\theta}\ne1,$ where $\theta\in(0,\pi],$ be a root of $p_{d,\alpha}$ on the unit circle. Writing $G^{-1}(e^{\jmath\theta}) = a + b\jmath,$ it follows that $a = \cos \theta + p$ and $b = \sin \theta,$ and \eqref{propAlphaTheta} and \eqref{propAlphaTheta2} have the form
\begin{align}
    \alpha(\theta) = \frac{e^{\jmath(d+2) \theta} + p e^{\jmath(d+1) \theta}}{e^{\jmath\theta}-1} = \frac{\cos \theta +p}{\cos d\theta - \cos\, (d+1)\theta},\label{alphaEqEx2_2}
\end{align}
which implies 
\begin{align}
    |\alpha(\theta)|=\sqrt{\frac{p^2 + 2p \cos \theta + 1}{2-2\cos\theta}}.
    \label{alphaAbsEqEx2_2}
\end{align}
Furthermore, it follows from \eqref{propbthetaa} that
\begin{align}
    \sin\, (d+2)\theta = (1-p) \sin\, (d+1)\theta + p \sin d \theta.\label{sinidentEx2_2}
\end{align}
Since $L_d$ has $d+2$ poles in the open unit disk and one zero at 1, it follows that there exist exactly $d+1$ distinct values $\theta_1,\ldots,\theta_{d+1}$ of $\theta \in [0,\pi]$ that satisfy \eqref{sinidentEx2_2}. 
%this isnt true-----how do we know the poles dont double back into OUD?
%
%This is because there are d+2 roots and ONE zero, so d+1 roots leave the unit disk
%
The corresponding values of $\alpha(\theta_i)$ are given by
\begin{align} \label{alphaEqEx2_2b}
    \alpha(\theta_i) &= \frac{\cos \theta_i +p}{\cos d\theta_i - \cos\, (d+1)\theta_i} \nn \\
    &= \frac{-\cos\,(d+2)\theta_i + (1-p)\cos\,(d+1)\theta_i + p \cos d\theta_i}{2 - 2 \cos \theta_i}.
\end{align}

Next, \textit{v}) in Theorem \ref{theorem_1} and \eqref{alphaAbsEqEx2_2} imply that
\begin{equation}\label{alphaInfEx2_2}
    \alpha_{\infty} = \min_{\theta\in(0,\pi]}\left|\frac{e^{\jmath\theta}+p}{e^{\jmath\theta}-1}\right| = \min_{\theta\in(0,\pi]} \sqrt{\frac{p^2 + 2p \cos \theta + 1}{2-2\cos\theta}}.
\end{equation}
Hence, it follows from \eqref{alphaAbsEqEx2_2} and \eqref{alphaInfEx2_2} that
\begin{equation}
    \alpha_{\infty} = \min_{\theta\in(0,\pi]} |\alpha(\theta)|.
\end{equation}
Letting $\theta^* \in (0, \pi]$ be a minimizer of \eqref{alphaAbsEqEx2_2}, it follows that
%
%
% \begin{equation*}
%     \left. \frac{d|\alpha|}{d\theta} \right|_{\theta = \theta^*} = 0,
% \end{equation*}
% %
%  implies   
%
\begin{equation}\label{dalphadThetaEx2_2}
    \left. \frac{d|\alpha|}{d\theta} \right|_{\theta = \theta^*} = -\frac{1}{2 |\alpha(\theta^*)|} \frac{\sin \theta^* (2p^2 + 4p + 2)}{(2-2 \cos \theta^*)^2} = 0,
\end{equation}
%
%It follows from \eqref{dalphadThetaEx2_2} and $\theta^* \in (0,\pi]$ that $\theta^* = \pi.$
%
which implies that $\theta^* = \pi.$
Hence, \eqref{alphaInfEx2_2} implies
\begin{equation}\label{alphainfValEx2_2}
    \alpha_{\infty} = \frac{1-p}{2} \in (0,1).
\end{equation}

%
% It can be shown that, as in the previous example, . It follows from \eqref{alphaAbsEqEx2_2} that the minimal value for $|\alpha (\theta)|$ satisfies
% %
% \begin{align}
%     \frac{\rm d |\alpha (\theta)|}{\rm d \theta} &= -\frac{\sin \theta (p^2 + 2p + 1)}{|\alpha (\theta)|(2 - 2 \cos \theta)} = 0, \label{alphaAbsdEq_Ex2_2}
% \end{align}
%It can be shown that for $\theta\in[0, \pi]$ and $p \in (-1, 1)$, only $\theta = \pi$ satisfies \eqref{alphaAbsdEq_Ex2_2}. Furthermore, since \eqref{alphaAbsdEq_Ex2_2} is negative for $\theta < \pi$ and positive for $\theta > \pi$, then the minimum magnitude for $\alpha(\theta)$ is $|\alpha(\pi]| = \alpha_{\infty} = \frac{|1-p|}{2}$. Note that $\theta = \pi$ satisfies \eqref{sinidentEx2_2} and that $\alpha(\pi) = (-1)^{d+1}\frac{1-p}{2}$.
%
%Furthermore, in the case where $d$ is odd, it follows that $\alpha_{d,\rml} = \frac{p-1}{2}<0$ and $\alpha_{d,\rmu} =  \alpha(\theta)>0$, whereas, in the case where $d$ is even, it follows that $\alpha_{d,\rml} = \alpha(\theta)<0$ and $\alpha_{d,\rmu} =\frac{1-p}{2}>0$.
%
%Note that, for $p = 0,$ these equations are equivalent to those given in Example \ref{ex_2_1}.
%
For $p = \half$, $d=6,$ and $d=7$, Figure \ref{fig:S2_ex_2_2_1} shows $\alpha(\theta_i)$ and $|\alpha(\theta_i)|$ versus $\theta_i$.  Note that, for both values of $d,$ the minimum value of $|\alpha(\theta_i)|$ is $\alpha_\infty = \tfrac{1}{4},$ as stated by \eqref{alphainfValEx2_2}, which occurs at $\theta = \pi$.
Finally, Figure \ref{fig:S2_ex_2_2_3} shows $\alpha_{d,\rml}$ and $\alpha_{d,\rmu}$ versus $d$ for $p = 0.5,$ which indicates that $\lim_{d\to\infty} -\alpha_{d,\rml} = \lim_{d\to\infty} \alpha_{d,\rmu} = \alpha_\infty,$ as stated in \eqref{alphainf}.
%
%Figure \ref{fig:S2_ex_2_2_1} shows $\alpha(\theta_k)$ versus $\theta_k$ and $p \in (-1, 1)$ for $d=10$ and $d=20$. 
%
%For the same values of $d$, Figure \ref{fig:S2_ex_2_2_2} shows $|\alpha(\theta_k)|$ versus $\theta_k$ and $p$.  Note that, for both values of $d,$ the minimum value of $|\alpha(\theta_k)|$ is 0.5, which occurs at $\theta = \pi$.
%
%Finally, Figure \ref{fig:S2_ex_2_2_3} shows $\alpha_{d,\rml}$ and $\alpha_{d,\rmu}$ versus $d$ and $p$.
%
%\hfill{\large$\diamond$}

\begin{figure}[h!]
    \centering
    \includegraphics[width=0.9\columnwidth]{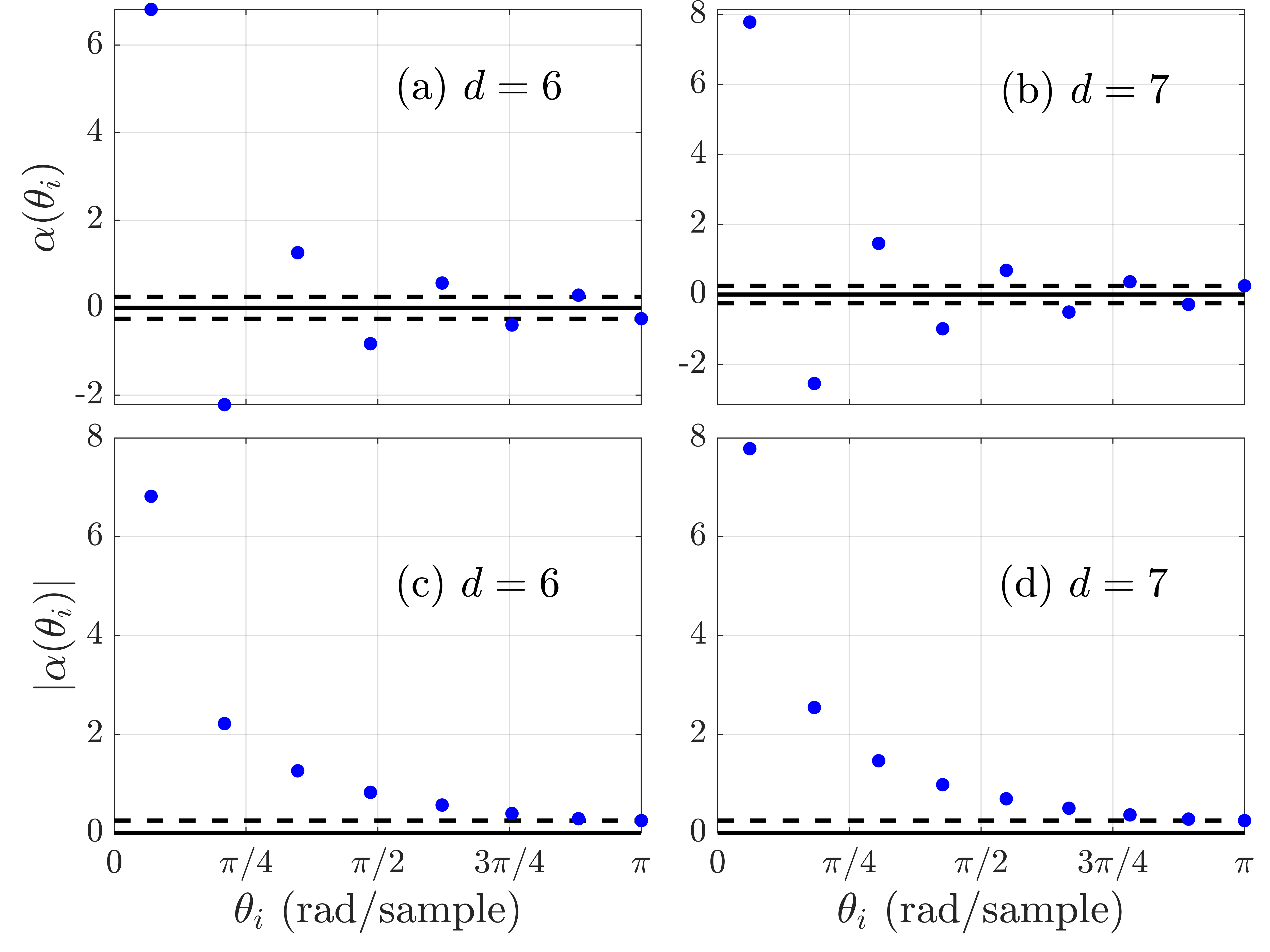} 
    \caption{\footnotesize Example \ref{ex_2_2}: For $p=\half$, $d=6,$ and $d=7,$ (a) and (b) show $\alpha(\theta_i)$ versus $\theta_i$, and (c) and (d) show $|\alpha(\theta_i)|$ versus $\theta_i$.  Note that the sign of $\alpha(\theta_i)$ alternates. The dashed lines indicate $\pm\alpha_\infty = \pm \tfrac{1}{4}.$ }
    \label{fig:S2_ex_2_2_1}
\end{figure}

\begin{figure}[h!]
    \centering
    \includegraphics[width=\columnwidth]{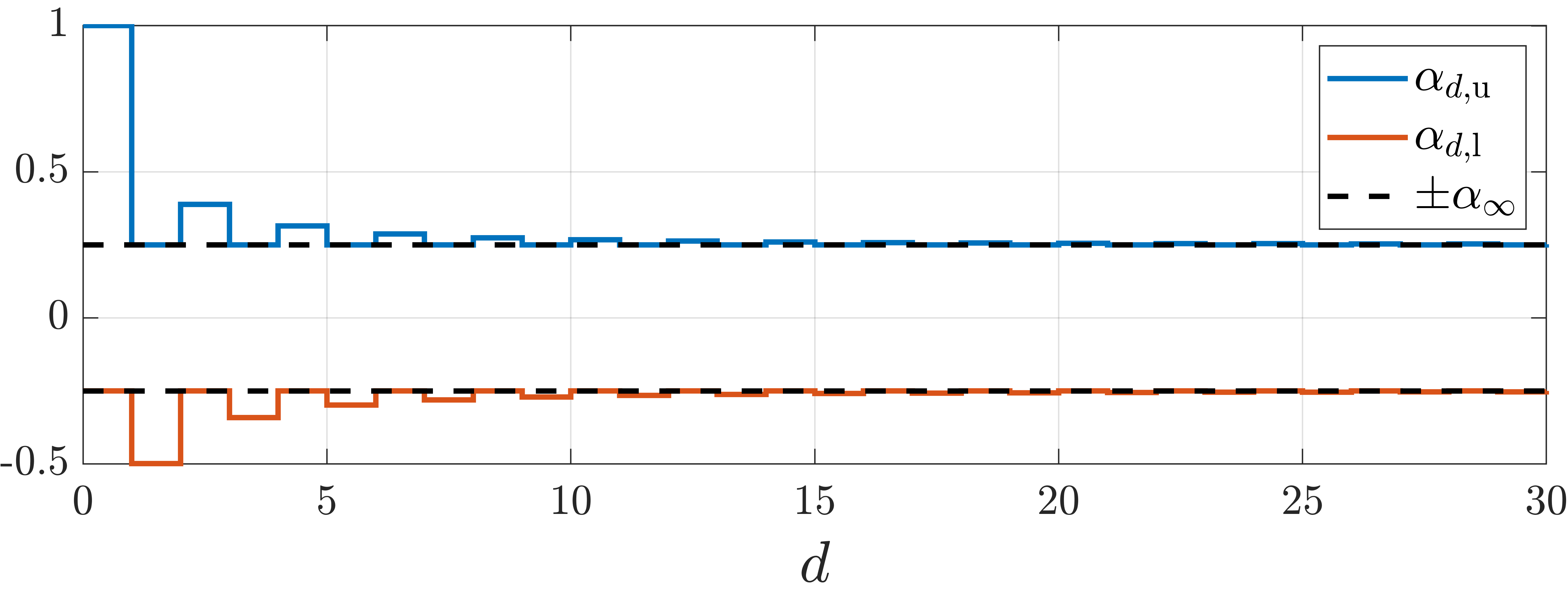}
    \caption{\footnotesize Example \ref{ex_2_2}: For $p = \half$, $\alpha_{d, {\rm l}}$ and  $\alpha_{d, {\rm u}}$ versus $d$.  As $d\to\infty,$ $\alpha_{d, {\rm l}}$ and  $\alpha_{d, {\rm u}}$ converge to $-\alpha_\infty$ and $\alpha_\infty$, respectively, where $\alpha_\infty = \tfrac{1}{4}.$}
    \label{fig:S2_ex_2_2_3}
\end{figure}
%%%SPECIAL CASE
\textbf{Special case:} For $p = 0,$ \eqref{alphaAbsEqEx2_2} becomes
\begin{align}
    |\alpha(\theta)|=\frac{1}{\sqrt{2-2\cos\theta}},
    \label{alphaAbsEq}
\end{align}
and \eqref{sinidentEx2_2} becomes
\begin{align}
    \sin\, (d+1)\theta = \sin\, (d+2)\theta. \label{sinident}
\end{align}
Note that, for all $i\in\BBZ,$ $\sin(d+1) \theta = \sin [(2i + 1)\pi - (d+1)\theta].$
Therefore, \eqref{sinident} holds if and only if $\theta = \frac{2k + 1}{2d + 3}\pi$.
Hence, $\theta\in[0, \pi]$ satisfies \eqref{sinident} if and only if  there exists $i\in\{0,\ldots,d+1\}$ such that $\theta_k\isdef \left(\frac{2i + 1}{2d + 3}\right)\pi.$
For these $d+2$ values of $\theta,$  \eqref{alphaEqEx2_2b} implies that the corresponding values of $\alpha(\theta)$ are given by 
\begin{align}
    \alpha(\theta_i) 
    &= \frac{ \cos \theta_i}{\cos d\theta_i - \cos (d+1)\theta_i}\nn\\[.5ex]
    &= \frac{\cos (d+1)\theta_i - \cos (d+2)\theta_i}{2 - 2 \cos \theta_i}.\label{alphaEqExp}
\end{align}
Next, it can be shown that, for all $i\in\{1,\ldots,d\},$ $\alpha(\theta_i) \alpha(\theta_{i+1}) < 0$.
Note that $\theta_{d+1} = \pi$ and $\alpha(\theta_{d+1})= (-1)^{d+1}\half$. Hence, $|\alpha(\theta_{d+1})| = \half.$
Furthermore, in the case where $d$ is even,
$\alpha_{d,\rml} = \alpha(\theta_{d+1}) = -\half<0$ and $\alpha_{d,\rmu} =  \alpha(\theta_d)>\half>0,$
whereas, in the case where $d$ is odd, $\alpha_{d,\rml} = \alpha(\theta_d)<-\half<0$ and $\alpha_{d,\rmu} = \alpha(\theta_{d+1}) = \half>0$.
In addition, although
$\lim_{d\to\infty}\alpha(\theta_d)$ does not exist, it follows from \eqref{alphaAbsEq} that
$\lim_{d \to \infty} |\alpha(\theta_d)| = \lim_{d \to \infty} \frac{1}{\sqrt{2-2\cos \left(\frac{2d + 1}{2d + 3}\right)\pi}} = \half$, which confirms \eqref{alphainf} and \eqref{alphainfValEx2_2}.
For $d=10$ and $d=11$, Figure \ref{fig:S2_ex_2_1_1} shows $\alpha(\theta_i)$ and $|\alpha(\theta_i)|$ versus $\theta_i$.  Note that, for both values of $d,$ the minimum value of $|\alpha(\theta_i)|$ is $\half$, which occurs at $\theta = \pi$.
Finally, Figure \ref{fig:S2_ex_2_1_3} shows $\alpha_{d,\rml}$ and $\alpha_{d,\rmu}$ versus $d,$ which indicates that $\lim_{d\to\infty} \alpha_{d,\rml} = -\half$ and $\lim_{d\to\infty} \alpha_{d,\rmu} = \half.$
\hfill{\large$\diamond$}

\begin{figure}[h!]
    \centering
     \includegraphics[width=0.9\columnwidth]{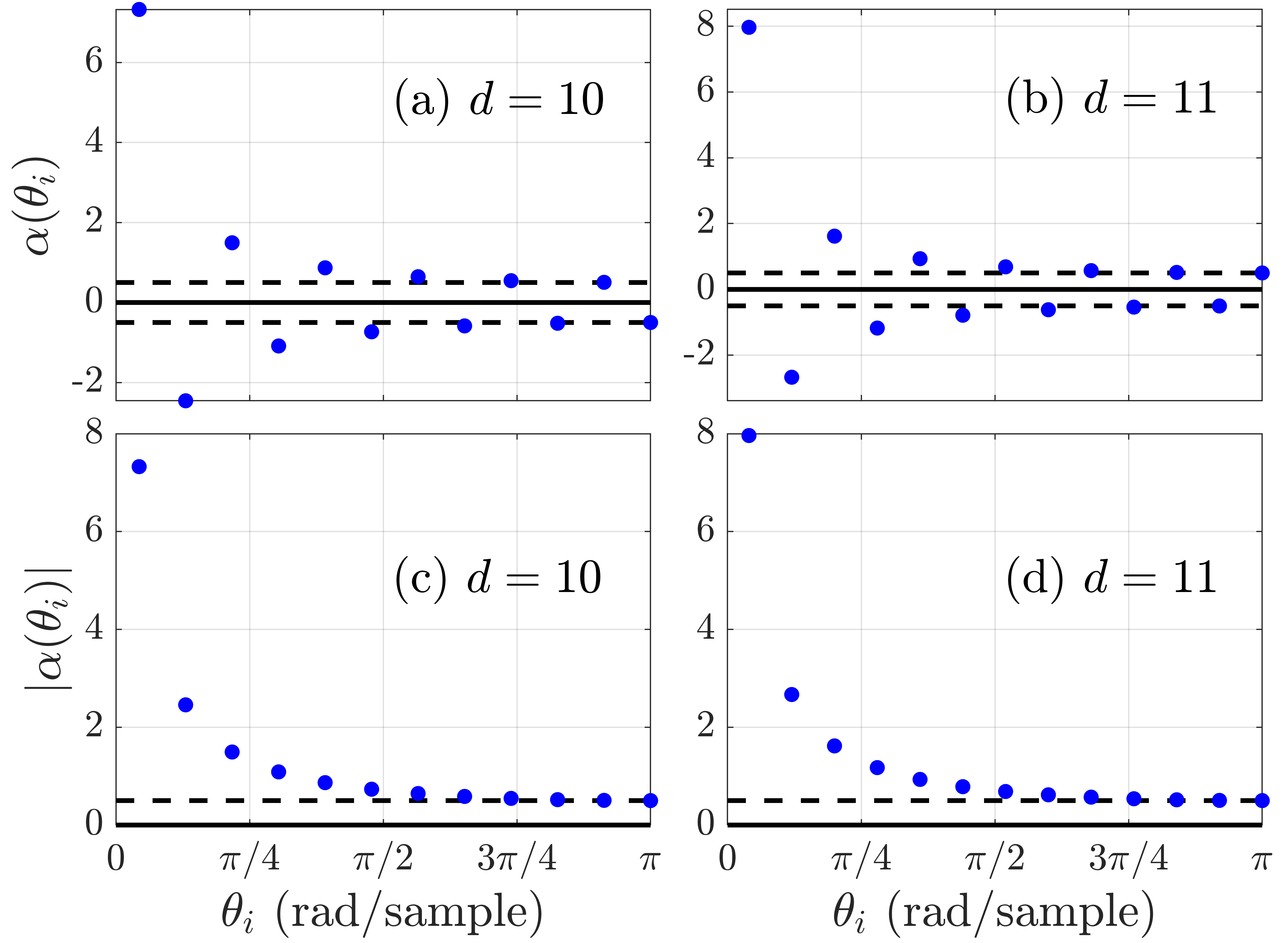} 
    \caption{\footnotesize Example \ref{ex_2_2}: For $p=0,$ $d=10,$ and $d=11,$ (a) and (b) show $\alpha(\theta_i)$ versus $\theta_i$, and (c) and (d) show $|\alpha(\theta_i)|$ versus $\theta_i$. The dashed lines indicate $\pm\alpha_\infty = \pm \half.$ 
    %Note that the signs of $\alpha(\theta_i)$ alternate.
    }
    \label{fig:S2_ex_2_1_1}
\end{figure}
% \begin{figure}[h]
%     \centering
%       \begin{subfigure}[h]{0.5\columnwidth}
%         \centering
%         \includegraphics[width=\columnwidth]{figures/Section_2/ex_2_1_2_d10.jpg} 
%         \caption{\footnotesize $d = 10$} 
%         \label{figS2:Ex2_1_2_d10}
%       \end{subfigure}%% 
%       \begin{subfigure}[h]{0.5\columnwidth}
%         \centering
%         \includegraphics[width=\columnwidth]{figures/Section_2/ex_2_1_2_d11.jpg} 
%         \caption{\footnotesize $d = 11$} 
%         \label{figS2:Ex2_1_2_d11}
%       \end{subfigure}
%     \caption{\footnotesize Example 2.1: $|\alpha(\theta_k)|$ versus $\theta_k$ .}
%     \label{fig:S2_ex_2_1_2}
% \end{figure}

\begin{figure}[h!]    
    \centering
    \includegraphics[width=\columnwidth]{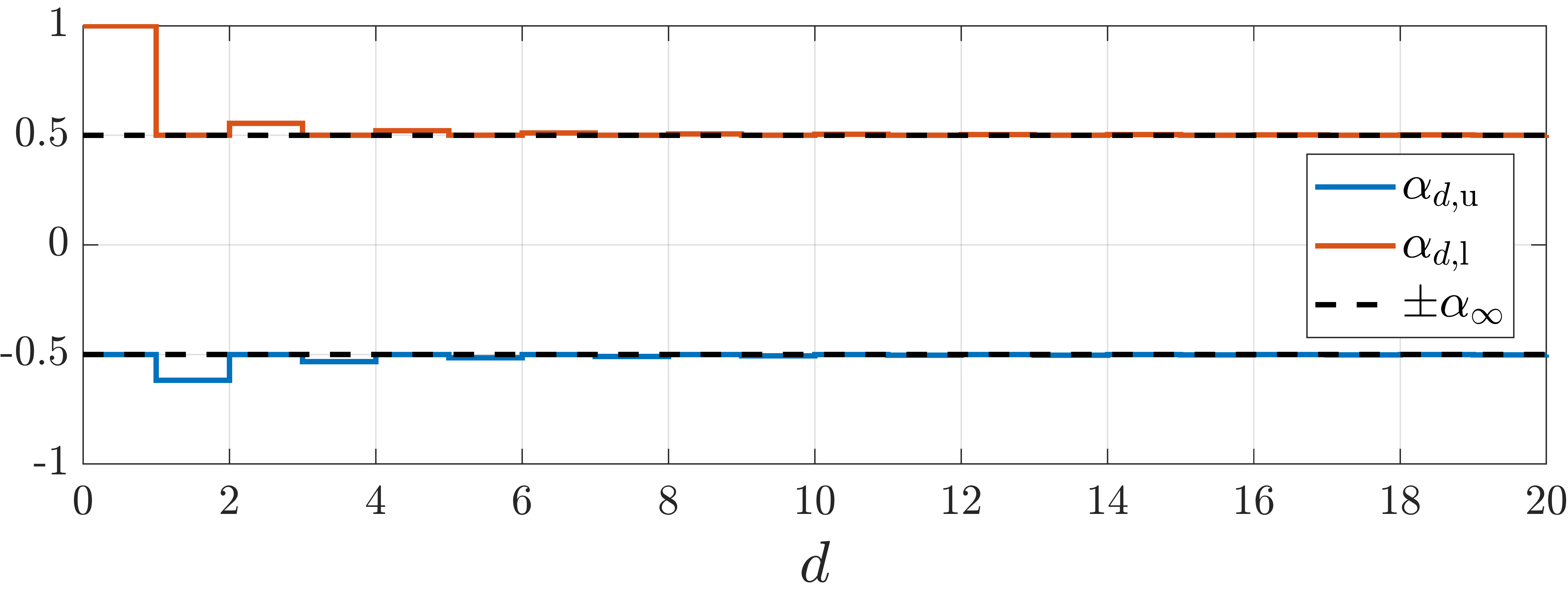}
     \caption{\footnotesize Example \ref{ex_2_2}:  For $p = 0$, $\alpha_{d, {\rm l}}$ and  $\alpha_{d, {\rm u}}$ versus $d$.  As $d\to\infty,$ $\alpha_{d, {\rm l}}$ and  $\alpha_{d, {\rm u}}$ converge to $-\alpha_\infty$ and $\alpha_\infty$, respectively, where $\alpha_\infty = \half.$}\label{fig:S2_ex_2_1_3}
\end{figure}
\end{exam}

\begin{exam} \label{ex_2_3}
Let
\begin{equation*}
    G(z) = \frac{N(z)}{D(z)} = \frac{z + 0.2 \pm \jmath 0.79}{z (z - 0.25 \pm \jmath 0.95)}.
\end{equation*}
Figure \ref{fig:S2_ex_2_3_1} shows that, for all $d \geq 1,$ there exists $\alpha_{d,\rml}<0$ such that $p_{d,\alpha}$ if and only if $\alpha \in (\alpha_{d, \rml},0]$ as stated in \textit{iv}) from Theorem \ref{theorem_1}.
Furthermore, define
\small
\begin{equation*}
    \alpha_{\rm uc}(\theta) \isdef \left|\frac{D(e^{\jmath\theta})}{(e^{\jmath\theta}-1)N(e^{\jmath\theta})}\right| = \left|\frac{(e^{\jmath\theta} - 0.25 \pm \jmath 0.95)}{(e^{\jmath\theta}-1)(e^{\jmath\theta} + 0.2 \pm \jmath 0.79)}\right|,
\end{equation*}
\normalsize
such that $\alpha_\infty = \min_{\theta\in(0,\pi]} \alpha_{\rm uc}(\theta).$
Figure \ref{fig:S2_ex_2_3_2} shows that $\alpha_{\rm uc}$ has a minimum at $\theta \approx 0.4180 \pi,$ which implies that $\alpha_\infty \approx \alpha_{\rm uc}(0.4180 \pi) = 0.0313.$
Finally, Figure \ref{fig:S2_ex_2_3_3} shows $\alpha_{d,\rml}$ and $\alpha_{d,\rmu}$ versus $d\geq 1,$ which shows that $\lim_{d\to\infty} \alpha_{d,\rml} = -\alpha_\infty$ and $\lim_{d\to\infty} \alpha_{d,\rmu} = \alpha_\infty,$ as stated in \textit{v}) from Theorem \ref{theorem_1}.
\hfill{\large$\diamond$}

\begin{figure}[h!]    
    \centering
    \includegraphics[width=\columnwidth]{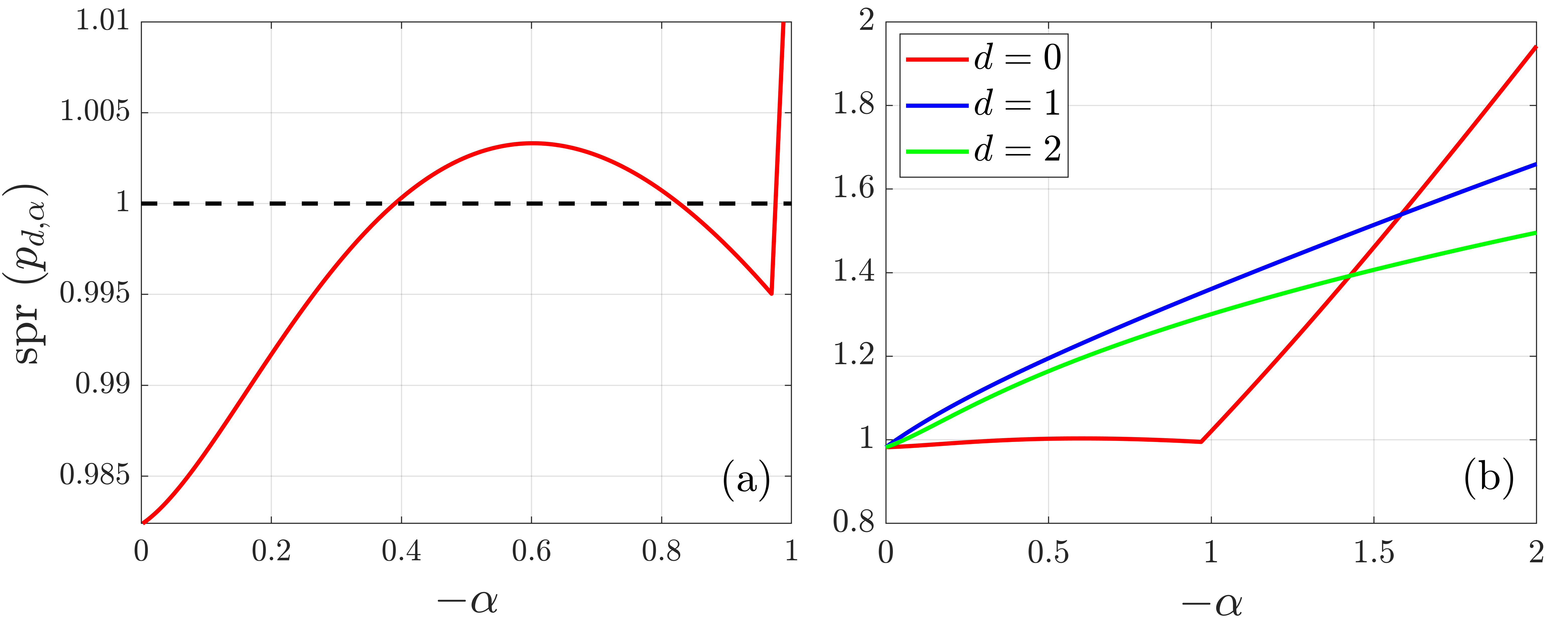}
     \caption{\footnotesize Example \ref{ex_2_3}: $\spr (p_{d,\alpha})$ versus $-\alpha,$ for $\alpha \leq 0.$ (a) shows that, for $d = 0,$ $\alpha_{0,\rml}$ as defined in \textit{iv}) from Theorem \ref{theorem_1} doesn't exist. (b) shows that, for $d\geq1,$ $\alpha_{0,\rml}$ as defined in \textit{iv}) from Theorem \ref{theorem_1} exists.}\label{fig:S2_ex_2_3_1}
\end{figure}
%
%\vspace{-2em}
%
\begin{figure}[h!]    
    \centering
    \includegraphics[width=\columnwidth]{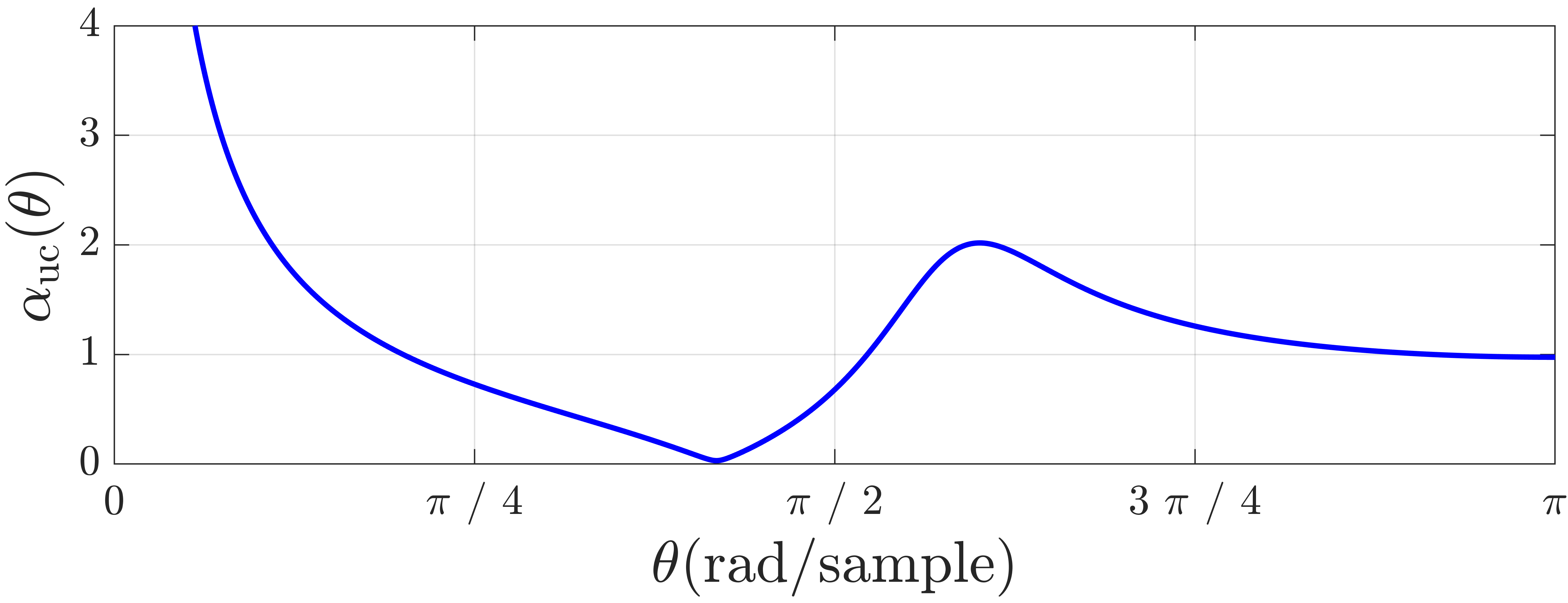}
     \caption{\footnotesize Example \ref{ex_2_3}: $\alpha_{\rm uc}(\theta)$ versus $\theta$ for $\theta \in (0,\pi].$ Note that $\lim_{\theta\downarrow0} \alpha_{\rm uc}(\theta) = \infty$ and that $\min_{\theta\in(0,\pi]} \alpha_{\rm uc}(\theta) = \alpha_{\rm uc}(\theta_\infty) \approx 0.0313,$ where $\theta_\infty \approx 0.4180 \pi.$}\label{fig:S2_ex_2_3_2}
\end{figure}
\begin{figure}[h!]    
    \centering
    \includegraphics[width=\columnwidth]{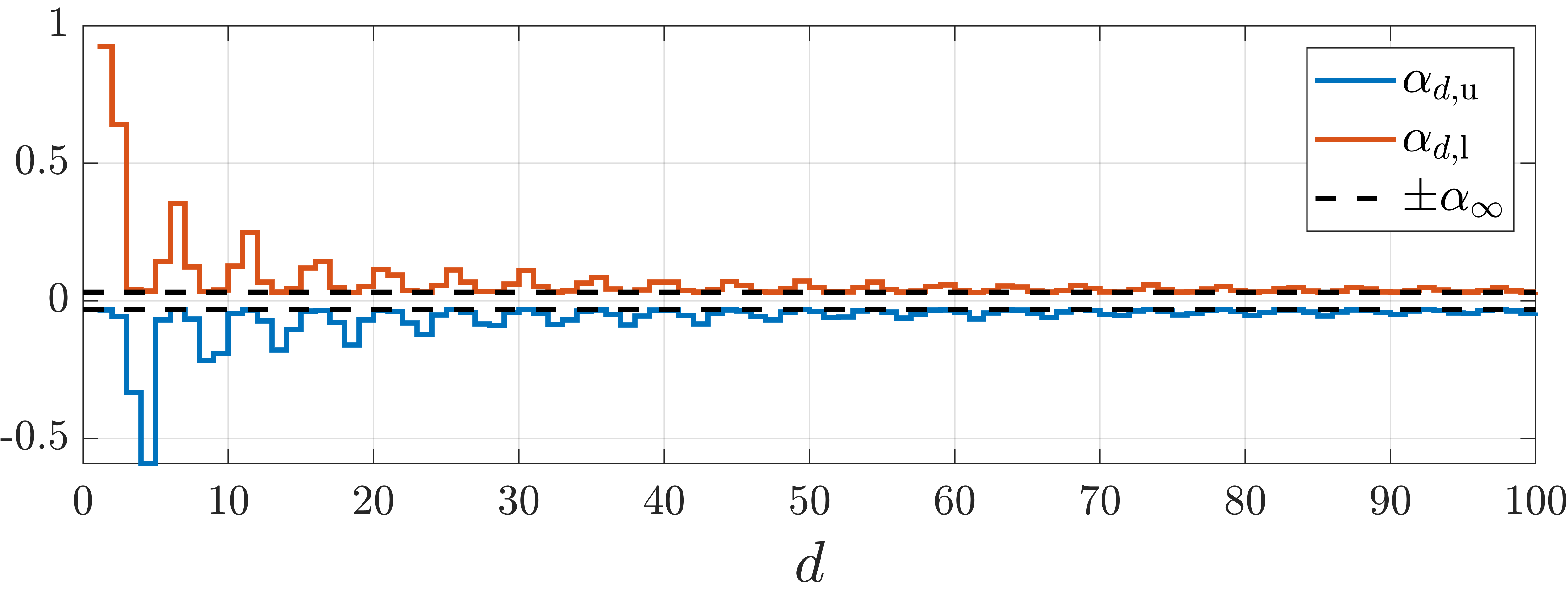}
     \caption{\footnotesize Example \ref{ex_2_3}: $\alpha_{d, {\rm l}}$ and  $\alpha_{d, {\rm u}}$ versus $d\geq 1.$  As $d\to\infty,$ $\alpha_{d, {\rm l}}$ and  $\alpha_{d, {\rm u}}$ converge to $-\alpha_\infty$ and $\alpha_\infty$, respectively, where $\alpha_\infty \approx 0.0313.$}\label{fig:S2_ex_2_3_3}
\end{figure}
\end{exam}

%\newpage

\section{Time-Delayed Lur'e Model}

%"locally not asymptotically stable, globally stabilized"

%show the effect of d on the spectrum of the oscillations

%show the effect of alpha on the oscillations----amplitude?

%show the effect of delta on amplitude?

\begin{figure}[h!]
    \centering
     \resizebox{0.75\columnwidth}{!}{%
    \begin{tikzpicture}[>={stealth'}, line width = 0.25mm]
    \node [input, name=input] {};
    \node [smallblock, rounded corners, right = 2cm of input, minimum height = 0.6cm, minimum width = 0.8cm] (system) {$\alpha  G(z)$};
    \draw [->] (input) -- node[name=usys] {} (system);
    \node [output, right = 2.2cm of system] (output) {};
    \node [smallblock, rounded corners, below = 0.3cm of system, minimum height = 0.6cm, minimum width = 0.8cm] (diff) {$W(z)$};
    \node [smallblock, rounded corners, right = 0.6cm of diff, minimum height = 0.6cm, minimum width = 0.8cm] (delay) {$G_d(z)$};
    %\node [saturation block, left = 0.6cm of diff, minimum width=1.25cm, minimum height=2.5em] (satq) {};
    \node [smallblock, rounded corners, left = 0.6cm of diff, minimum height = 0.6cm, minimum width = 0.8cm](satq){${\rm sat}_\delta$};
    \draw [-] (satq.west) -| node [near start, above] {$v_{\rm f}$} (input);
    \draw [->] (system) -- node [name=y, very near end]{} node [very near end, above] {$y$}(output);
    \draw [->] (y.center) |- (delay);
    \draw [->] (delay) -- node [above] {$y_\rmd$} (diff);
    \draw [->] (diff) -- node [above] {$y_{\rm f}$} (satq);
    \end{tikzpicture}
    }
    \caption{\footnotesize Discrete-time time-delayed Lur'e model.}
    \label{DT_TDL_wSat_blk_diag}
\end{figure}
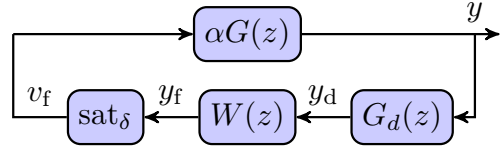

Inserting the saturation nonlinearity following the washout filter $W$ in Figure \ref{DT_TDL_woSat_blk_diag} yields the
%
%{\it  time-delayed Lur'e} (TDL)
%time-delayed Lur'e
TDL
model shown in Figure \ref{DT_TDL_wSat_blk_diag}, which has the closed-loop dynamics
\begin{align} 
    \left[ \arraycolsep=1.1pt\def\arraystretch{1.2} \begin{array}{c} x_{k+1}\\ x_{\rmd,k+1}\\ x_{\rmf,k+1} \end{array}\right]
    = \left[ \arraycolsep=1.6pt\def\arraystretch{1.2} \begin{array}{ccc} A & 0 & 0\\ 
    e_{d,d}C & N_{d} & 0\\ 0 & e_{1,d}^\rmT   & 0  \end{array}\right]
    \left[ \arraycolsep=1.1pt\def\arraystretch{1.2} \begin{array}{c} x_{k}\\ x_{\rmd,k}\\ x_{\rmf,k}   \end{array}\right]
     +  \left[ \arraycolsep=1.1pt\def\arraystretch{1.2} \begin{array}{c} \alpha B\\ 0 \\ 0 \end{array}\right] v_{{\rm f}, k}, \label{TDLeqn}
\end{align}
\normalsize
with $y_k,$ $y_{\rmd,k},$ and $y_{\rmf,k}$  given by \eqref{TDLeqnlin}, \eqref{ydk}, and \eqref{yfk}, respectively,  where $v_{{\rm f},k} = {\rm sat}_\delta (y_{{\rm f},k})$ is the output of the saturation function ${\rm sat}_\delta \colon\BBR\to\BBR$, where $\delta > 0,$ defined by
\begin{align}
    {\rm sat}_\delta (u) =
    \begin{cases}
    u, & |u| \le \delta,\\
    {\rm sign}(u), & |u| > \delta.
    \end{cases}
\end{align}
%
%
%in the linear feedback system as shown in Figure \ref{DT_TDL_wSat_blk_diag} yields the {\it  time-delayed Lur'e} (TDL) system.

% 
To analyze the self-oscillating behavior of the time-delayed Lur'e model, we replace the saturation nonlinearity by its describing function.
Describing functions are used to characterize self-excited oscillations in \cite[Section 5.4]{Ding2010} and \cite[pp. 293--294]{khalil3rd}.
The describing function  $\Psi_\delta (\varepsilon)$ for $\rm sat_{\delta}$ for a sinusoidal input with amplitude $\varepsilon > 0$ is given by
\begin{align}
    \Psi_\delta (\varepsilon) &=
    \begin{cases}
    \frac{2}{\pi} \left[\sin^{-1} \left(\frac{\delta}{\varepsilon}\right) + \left(\frac{\delta}{\varepsilon}\right) \sqrt{1 - \left(\frac{\delta}{\varepsilon}\right)^2} \right],& \text{if } \varepsilon > \delta\\
    1,              & \text{otherwise}.
    \end{cases}
\end{align}
Note that, for $\varepsilon > \delta$, the function $\Psi_\delta$ confined to $(\delta,\infty)$ with codomain $(0,1)$ is decreasing, one-to-one, and onto.
Let $p_{d, \alpha, \varepsilon}$ be the characteristic polynomial of the linearized time-delay Lur'e model, such that
\begin{align}
    p_{d,\alpha,\varepsilon}(z)\isdef z^{d+1} D(z) - \alpha \Psi_\delta(\varepsilon) (z-1)N(z).
\end{align}
For all $\varepsilon_\rml>0,$ $\varepsilon_\rmu>0,$  
$\theta_\rml\in\BBR,$ and $\theta_\rmu\in\BBR$ such that $\varepsilon_\rml<\varepsilon_\rmu$ and $\theta_\rml<\theta_\rmu,$
define the rectangle
\begin{align}
    \Gamma_{\theta_\rml,\theta_\rmu, \varepsilon_\rml, \varepsilon_\rmu} \isdef \{(\theta,\varepsilon)\colon 
    \theta_\rml<\theta<\theta_\rmu \mbox{ and }  
    \varepsilon_\rml<\varepsilon<\varepsilon_\rmu 
    \}\nn
\end{align}
%
% 
%JJJJJ: One of the conditions stated in the Khalil is that there exists a region around $(\theta, \varepsilon)$ in which this set is the unique solution to $p_{d,\alpha,\varepsilon_0}(e^{\jmath \theta_0}) = 0$. Khalil uses notation very freely to state this, so I didn't know how to state this with mathematical precision.
%
%
%
%
%
%

\begin{lem}
\label{lemma_3_1}
Let $\alpha \in \mathbb{R},$
and let $\theta_0\in\Theta$ be such that $\sign \alpha_0 = \sign \alpha$ and $|\alpha_0| < |\alpha|$, where 
$\alpha_0 \isdef  \alpha(\theta_0),$ and let $d > \bar{d}$.
Then, the following statements hold:
\begin{enumerate}
\item There exist $\varepsilon_0 > 0$,
$\theta_\rml>0,$ $\theta_\rmu>0,$ 
$\varepsilon_\rml>0,$ and $\varepsilon_\rmu>0$ 
such that
$\varepsilon_\rml<\varepsilon_\rmu,$  $\theta_\rml<\theta_\rmu,$
$(\theta_0,\varepsilon_0)\in \Gamma_{\theta_\rml,\theta_\rmu, \varepsilon_\rml, \varepsilon_\rmu},$
and, in the rectangle  $\Gamma_{\theta_\rml,\theta_\rmu, \varepsilon_\rml, \varepsilon_\rmu}$,
$(\theta,\varepsilon)=(\theta_0, \varepsilon_0)$ is the unique solution  of $p_{d,\alpha,\varepsilon}(e^{\jmath \theta}) = 0$.
\item
\begin{align}
    \left. \frac{\rmd}{\rmd \varepsilon} \Psi_\delta(\varepsilon) \right\rvert_{\varepsilon = \varepsilon_0} \neq 0.
\end{align}
\item
\begin{align}
    \left. \frac{\rmd}{\rmd \theta} {\rm Im}[L_d(e^{\jmath \theta})] \right\rvert_{\theta = \theta_0} \neq 0.
\end{align}
\end{enumerate}
\end{lem}

{\bf Proof.}
%\begin{proof}
%
To prove $i),$ note that, for $\sign \alpha_0 = \sign \alpha$ and $|\alpha_0| > |\alpha|$, there exists $\varepsilon_0 > \delta$ such that $\alpha_0 = \Psi_\delta(\varepsilon_0) \alpha$. Therefore, $p_{d, \alpha_0} (e^{\jmath \theta_0}) = p_{d, \alpha, \varepsilon_0} (e^{\jmath \theta_0}) = 0.$ %Furthermore, there exists 
%
% since all members of $\Theta$ are unique,?????
%
%
%
%
%$\Gamma_{\theta_0, \varepsilon_0}$ such that $(\theta_0, \varepsilon_0)$ is the unique solution in $\Gamma_{\theta_0, \varepsilon_0}$ for $p_{d,\alpha,\varepsilon_0}(e^{\jmath \theta_0}) = 0.$
%
Furthermore, there exists a rectangle $\Gamma_{\theta_\rml,\theta_\rmu, \varepsilon_\rml, \varepsilon_\rmu},$
 where $\theta_\rml>0,$ $\theta_\rmu>0,$ 
$\varepsilon_\rml>0,$ $\varepsilon_\rmu>0,$
$\varepsilon_\rml<\varepsilon_\rmu$ and $\theta_\rml<\theta_\rmu,$
such that $(\theta_0,\varepsilon_0)\in \Gamma_{\theta_\rml,\theta_\rmu, \varepsilon_\rml, \varepsilon_\rmu}$ and $\Theta \cap (\theta_\rml, \theta_\rmu) = \theta_0.$ Hence, in the rectangle  $\Gamma_{\theta_\rml,\theta_\rmu, \varepsilon_\rml, \varepsilon_\rmu}$,
$(\theta,\varepsilon)=(\theta_0, \varepsilon_0)$ is the unique solution  of $p_{d,\alpha,\varepsilon}(e^{\jmath \theta}) = 0$.

To prove $ii),$ note that
%, for $\varepsilon > \delta$, 
\begin{align}
    \left. \frac{\rmd}{\rmd \varepsilon} \Psi_\delta(\varepsilon) \right\rvert_{\varepsilon = \varepsilon_0} = -\frac{4 \delta \sqrt{\varepsilon_0^2 - \delta^2}}{\pi \varepsilon_0^3}<0.
\end{align}
% Since $\varepsilon_0 > \delta$, it follows that $ \left \frac{\rmd}{\rmd \varepsilon} \Psi_\delta(\varepsilon) \right\rvert_{\varepsilon = \varepsilon_k} \neq 0$.
%

To prove $iii)$, writing $G(e^{\jmath \theta}) = a_n-\jmath b_n,$ where $a_n=\frac{a}{a^2 + b^2}$ and $b_n = \frac{b}{a^2 + b^2}.$ Then,
\small
\begin{align} \label{eqLdFG}
    L_d&(e^{\jmath \theta}) = G(e^{\jmath \theta}) G_d(e^{\jmath \theta}) W(e^{\jmath \theta}) \nn \\
    &= (a_n - \jmath b_n) (e^{-\jmath d \theta} - e^{-\jmath (d + 1) \theta}) \nn \\
    &= (a_n (\cos d \theta - \cos\,(d+1) \theta) - b_n (\sin d \theta - \sin\, (d+1) \theta)) \nn \\
    & + \jmath (- a_n (\sin d \theta - \sin\, (d+1) \theta) - b_n (\cos d \theta - \cos\,(d+1) \theta)) \nn \\
    &= \frac{f}{a^2 + b^2} - \jmath \frac{g}{a^2 + b^2}.
\end{align}
\normalsize
%
%
%
%
%%%%%%%%%%%%%%%%%%%
Since $\theta_0 \in \Theta$ and $\alpha(\theta_0) \neq 0$, it follows from \eqref{propbthetaa} and \eqref{eqLdFG} with $\theta=\theta_0$ that $g = 0$ and thus
%
%Since $\theta_0 \in \Theta$ and $\alpha(\theta_0) \neq 0$, it follows from \eqref{alphaThetafg} that
%do you really mean (21) here?  (21) is just the defn of g, right?
%here
%ok.  From here to the end of the proof is very confusing.  Please try to write it extremely clearly in every step.  There are many parts real, imag, trig, angles, etc, so it is hard to follow the steps precisely.  Pls try to fix and we will go over it tomorrow.  OK?  thanks  also when you say something, try to give the reason such as why d > dbar.  Thanks  see you tomorrow, thanks
%
%
\begin{align} \label{anGLTheta_0_ReIm}
{\rm Re}[L_d(e^{\jmath \theta_0})] \neq 0, \quad
{\rm Im}[L_d(e^{\jmath \theta_0})] = 0.
\end{align}
%
%PLEASE READ!!!!!!!!!!!!!!!!!!!!!!!
%The real and imaginary expressions are similar to the equations in \eqref{alphaEqn1} for f and g. The above expression follows from the fact that f \neq 0 and g = 0 for $theta_0 \in Theta$. Is this the correct way of expressing this?
%
Furthermore, differentiating $\angle L_d(e^{\jmath \theta})$ with respect to $\theta$ yields 
\begin{align} \label{anGL_dTheta}
    & \frac{\rmd}{\rmd \theta} \angle L_d(e^{\jmath \theta}) = \frac{\rmd}{\rmd \theta} {\rm atan} \left(\frac{{\rm Im} [L_d(e^{\jmath \theta})]}{{\rm Re} [L_d(e^{\jmath \theta})]}\right) \nn \\
    & = \frac{{\rm Re} [L_d(e^{\jmath \theta})] \frac{\rmd}{\rmd \theta} {\rm Im}[L_d(e^{\jmath \theta})] - {\rm Im}[L_d(e^{\jmath \theta})] \frac{\rmd}{\rmd \theta} {\rm Re}[L_d(e^{\jmath \theta})]}{|L_d(e^{\jmath \theta})|}.
\end{align}
It follows from \eqref{anGLTheta_0_ReIm} and \eqref{anGL_dTheta} that
\begin{align} \label{Lejtheta_dtheta}
    \left.\frac{\rmd}{\rmd \theta} \angle L_d(e^{\jmath \theta}) \right\rvert_{\theta = \theta_0} = \frac{{\rm Re} [L_d(e^{\jmath \theta_0})] \left. \frac{\rmd}{\rmd \theta} {\rm Im}[L_d(e^{\jmath \theta})]\right\rvert_{\theta = \theta_0}}{|L_d(e^{\jmath \theta_0})|}.
\end{align}
%
%Juan:
%\left. is needed if you want \right but do not need the \left one.  You were missing the period after \left
%
%
%
%Since, for $d > \bar{d}$, $|L(e^{\jmath \theta_0})| < \infty$ and $\left\frac{\rmd}{\rmd \theta} \angle L(e^{\jmath \theta}) \right\rvert_{\theta = \theta_0} < 0$ it follows that $ \left \frac{\rmd}{\rmd \theta} {\rm Im}[L(e^{\jmath \theta})] \right\rvert_{\theta = \theta_0} \neq 0.$
%
It follows from \eqref{angLTheta_dtheta_leq0} that, for all $d > \bar{d},$ $\left. \frac{\rmd}{\rmd \theta} \angle L_d(e^{\jmath \theta}) \right\rvert_{\theta = \theta_0} < 0.$ Hence, it follows from \eqref{Lejtheta_dtheta} that $ \left. \frac{\rmd}{\rmd \theta} {\rm Im}[L_d(e^{\jmath \theta})] \right\rvert_{\theta = \theta_0} \neq 0.$
%
%\end{proof}
\hfill{$\square$}

%\textbf{Theorem 2.} 
\begin{theo}
\label{theorem_2}
Consider the discrete-time time-delayed Lur'e model  in Figure \ref{DT_TDL_wSat_blk_diag}, assume that $x_0\ne0,$ and let $\alpha\in(-\infty,\alpha_{d,\rml})\cup(\alpha_{d,\rmu},\infty).$ 
Then, there exists a nonconstant periodic function $\tau\colon\BBN\to\BBR$ such that $\lim_{k\to\infty}|y_k-\tau_k| = 0.$

{\bf Proof.}
Lemma \ref{lemma_3_1} implies that the  assumptions of Theorem 7.4 in \cite[pp. 293, 294]{khalil3rd} are satisfied.
It thus follows that the response is asymptotically periodic.   \hfill{$\square$}

It can be seen that Theorem \ref{theorem_2} holds in the case where the saturation function is replaced by an odd sigmoidal nonlinearity such as atan or tanh.

%\textbf{Proof:} $i$)   Writing $G = N/D$, where $D$ is monic, it follows that the characteristic polynomial $p$ of \eqref{TDLeqnlin} has the form
%\begin{align}
%    p(z) = {z}^{d+1} D(z) - \alpha (z+1) N(z).\label{charpol}
%\end{align}
%
%Since $p$ is monic and has degree $n+d+1,$ it follows from a root locus argument that at least one root of $p$ diverges along an asymptote.  Hence, for all sufficiently large values of $|\alpha|,$ $p$ has at least one root located outside the closed unit disk. This proves {\it i}).

%\textit{ii)} It will be assumed that the input to the saturation has a harmonic component. 
 \end{theo}
%
%The following statement will need some work. I am not sure how to state the fact that the frequency of the magnitude peaks of the frequency response of y will be determined by $\Theta$.
%maybe this is enough

%Furthermore, Theorem 7.4 implies that

\begin{figure}[h!]
    \centering
    \includegraphics[width=\columnwidth]{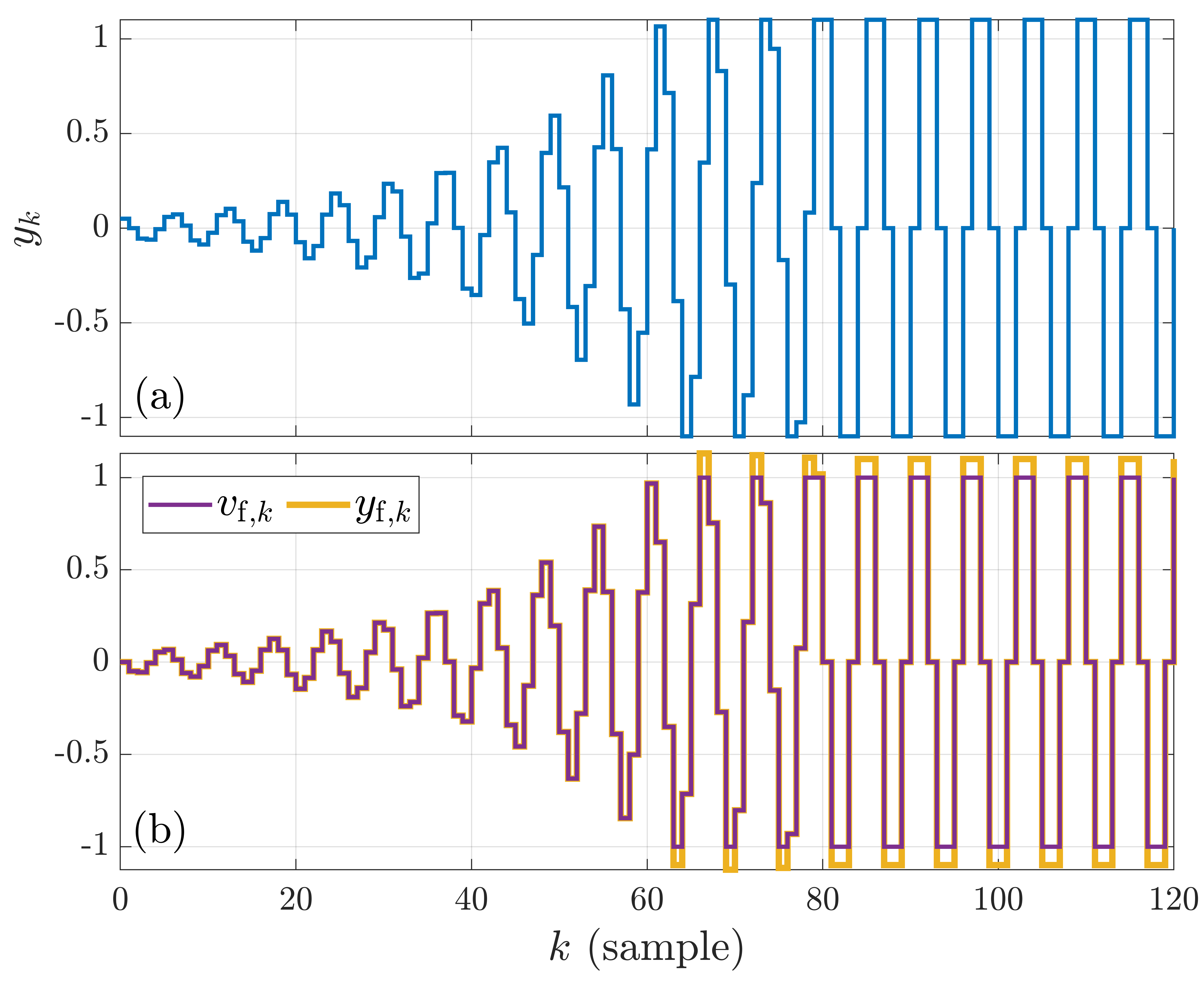}
    \caption{\footnotesize Example \ref{ex_3_1}: For $d = 0$, $\delta = 1,$ and $\alpha = 1.1,$ (a) shows $y_k,$ and (b) shows $v_{{\rm f},k}$ and $y_{{\rm f}, k}$.
    The saturation nonlinearity, with $\delta = 1,$ saturates the values of $y_{{\rm f}, k},$ resulting in $v_{{\rm f},k} \in [-\delta, \delta].$
    %
    %Note that the values of $v_{{\rm f}, k}$ alternate between the linear and saturated regions of the saturation function, which results in the oscillatory response of $y_k.$
    %Note that the amplitude of the oscillation is greater than $\delta = 1$ due to the fact that $y_k$ leaves and returns to the linear region of the saturation function.
    %
    }
    \label{fig:S3_ex_3_1_1}
\end{figure}

\begin{figure}[h!]
    \centering
      \includegraphics[width=\columnwidth]{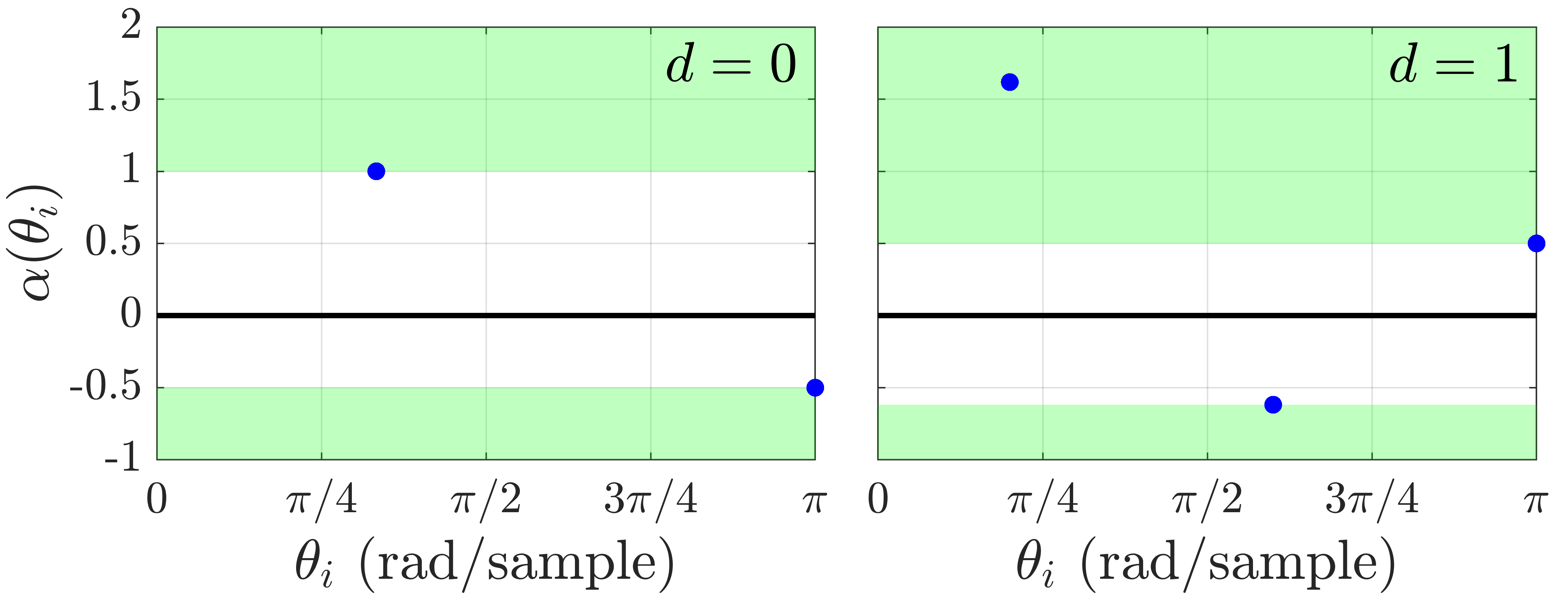}
    \caption{\footnotesize Example \ref{ex_3_1}: For $d = 0$ and $d = 1$, these plots show the values $\alpha(\theta_i)$ of $\alpha$ for which the closed-loop dynamics have a pole on the unit circle at the angle $\theta_i$ for $i=1,\ldots,d+2.$ %
    For the case $d=0,$ where $\theta_1 = 1.0472$ and $\theta_2 = \pi$, the time-delayed Lur'e model has self-excited oscillations if and only if either $\alpha>1$ or $\alpha < -\half,$ while, for the case $d=1,$ where $\theta_1 = 0.6283$, $\theta_2 = 1.8850$ and $\theta_3 = \pi$, the TDL model has self-excited oscillations if and only if either $\alpha>\half$ or $\alpha < -0.618.$ For all values of $\alpha$ corresponding to the shaded regions, the response of the TDL model oscillates.
    %
    %
    %The dashed lines correspond to $\alpha = 0.6$ and $\alpha = 1.1$. 
    }
    \label{fig:S3_ex_3_1_2}
\end{figure}

\begin{figure}[h!]
    \centering
    \includegraphics[width=\columnwidth]{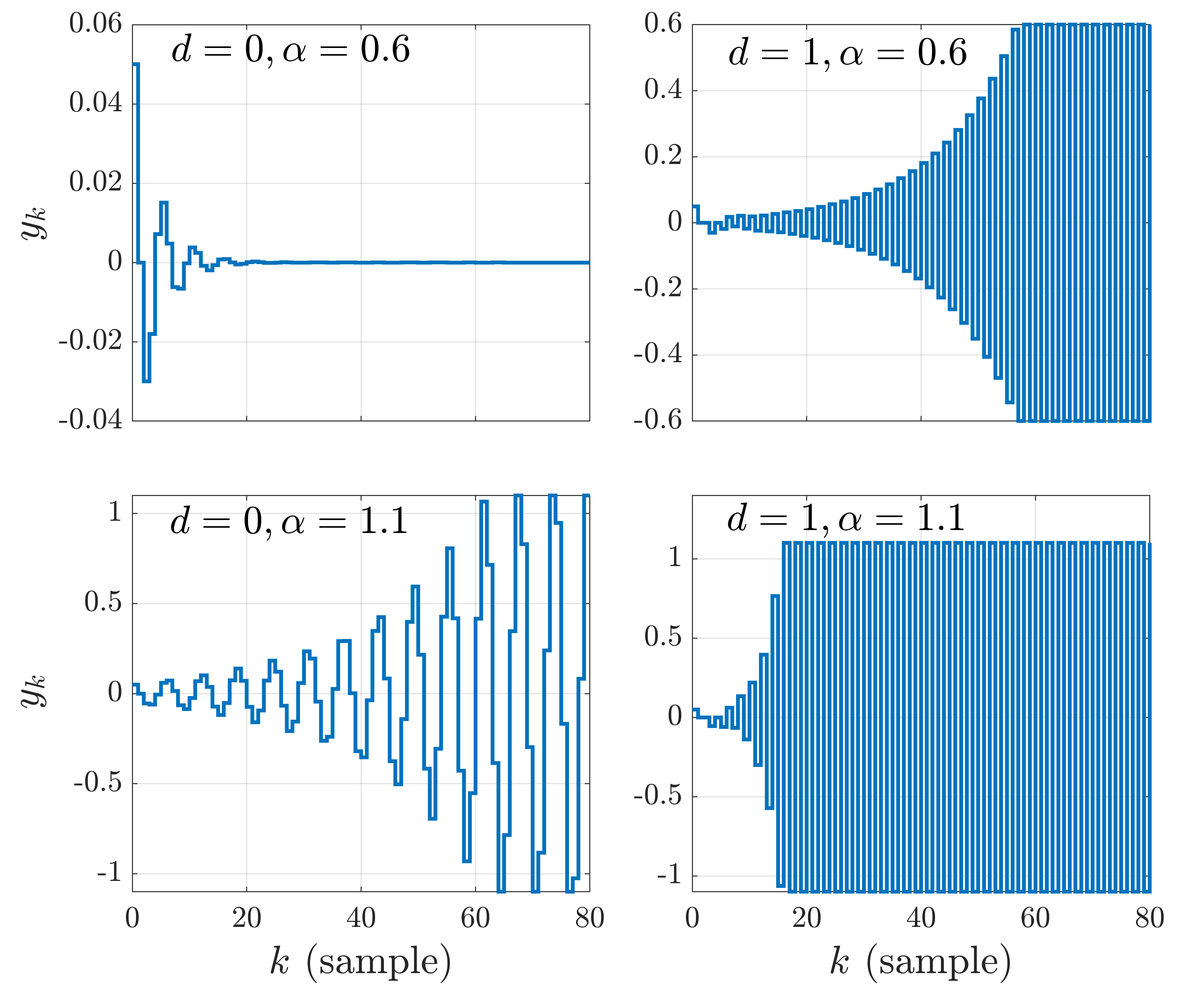}
    \caption{\footnotesize Example \ref{ex_3_1}: Response $y_k$  of the TDL model  for $d = 0, 1$ and $\alpha = 0.6, 1.1$ with $\delta = 1$.}
    \label{fig:S3_ex_3_1_3}
\end{figure}

\begin{figure}[h!]
    \centering
    \includegraphics[width=\columnwidth]{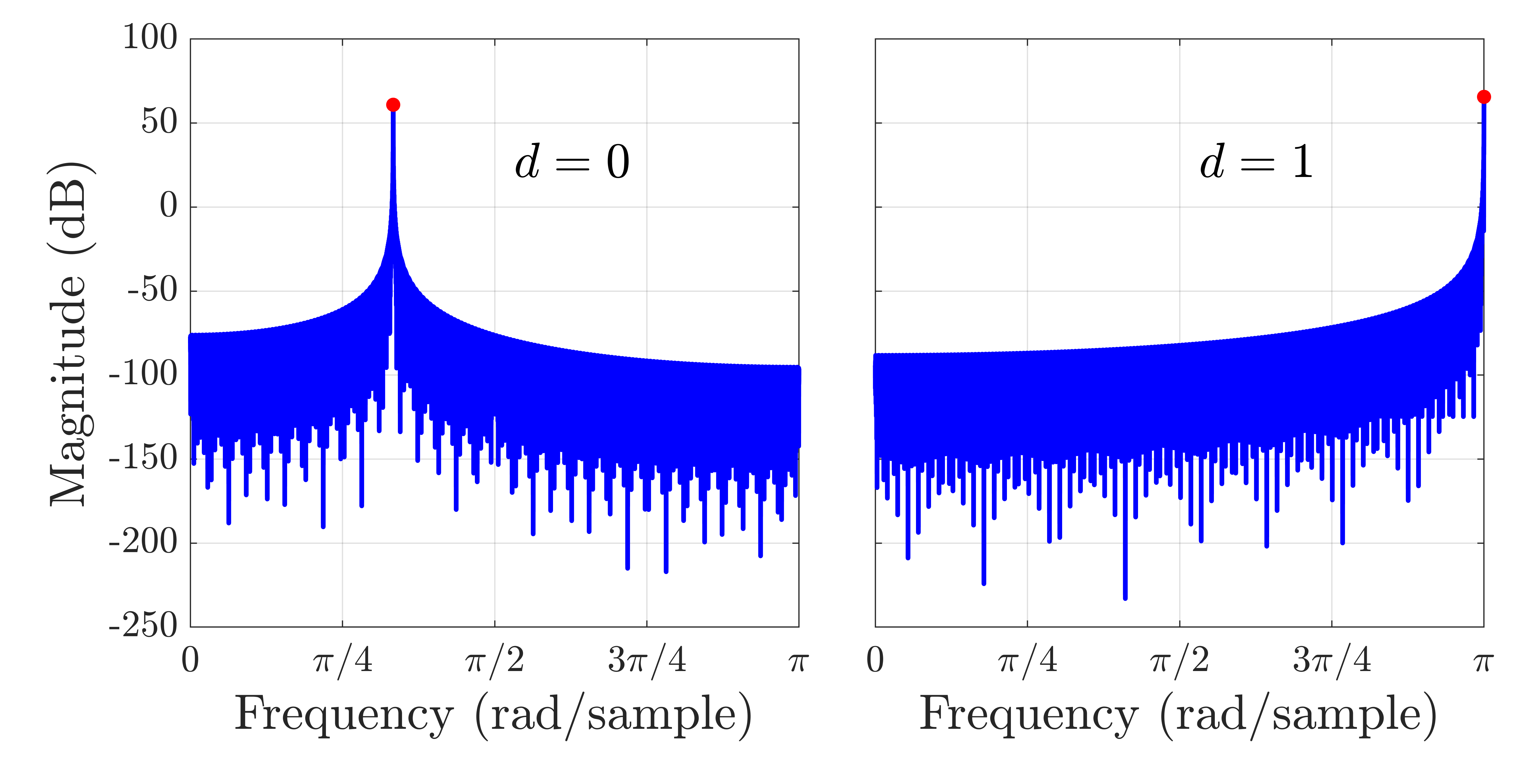}
    \caption{\footnotesize \footnotesize Example \ref{ex_3_1}: Frequency response of $y_k$ for $d = 0, 1$ with $\alpha = 1.1$ and $\delta = 1$. Note that, for $d=0,$ the peak is located at $\theta_1$, whereas, for $d=1,$ the peak is located at $\theta_3.$}
    \label{fig:S3_ex_3_1_4}\vspace{-.2in}
\end{figure}

%
%\textbf{Example 3.1.}
\begin{exam}
\label{ex_3_1}
Let $G(z)\mspace{-4mu} =\mspace{-5mu} 1/z.$ Figure \ref{fig:S3_ex_3_1_1} shows the transient response and  asymptotic oscillatory response for $\alpha = 1.1$, $d = 0$, and $\delta = 1$ along with plot of $v_{{\rm f}, k}$ and $y_{{\rm f}, k}$. 
Figure \ref{fig:S3_ex_3_1_1}(a) shows that, for $k>80$, $y_k$ is a nonconstant periodic function.
Furthermore, Figure \ref{fig:S3_ex_3_1_1}(b) shows how the saturation nonlinearity acts upon $y_{{\rm f}, k},$ which results in the saturated signal $v_{{\rm f}, k} \in [-\delta, \delta].$ Note that $v_{{\rm f}, k}$ and $y_{{\rm f}, k}$ are also nonconstant periodic functions for $k>80.$
%
%Furthermore, Figure \ref{fig:S3_ex_3_1_1}(b) shows how the model alternates between not asymptotically stable and asymptotically stable regions corresponding to the linear and constant segments of the saturation function.
%

%The role of $\alpha$ on the existence of oscillations will be shown next for $d = 0$ and $d = 1$.

Figure \ref{fig:S3_ex_3_1_2} shows $\alpha(\theta_i)$ versus $\theta_i$ for $d = 0$ and $d = 1$. For $\alpha = 0.6$, only in the case $d = 1$ has  $\alpha(\theta_i)$ such that $\rm sign (\alpha (\theta_i)) = \rm sign (\alpha)$ and $|\alpha (\theta_i)| < |\alpha|$. For $\alpha = 1.1$, both models meet the conditions for $\alpha$.

Figure \ref{fig:S3_ex_3_1_3} shows the response of $y_k$ for $\delta = 1$ and all possible pairs of $d = 0, 1$ and $\alpha = 0.6, 1.1$. For $\alpha = 0.6$, only the model with $d = 1$ yields a limit cycle. For $\alpha = 1.1$, both models yield oscillations. This follows from the conditions for $\alpha$ stated in the previous paragraph and in Lemma \ref{lemma_3_1}.

Finally, Figure \ref{fig:S3_ex_3_1_4} shows the magnitude of the frequency response for models with $\alpha = 1.1,$ $\delta = 1,$ and $d = 0, 1.$ Note that the frequencies corresponding to the magnitude peaks are similar to the values of $\theta_i$ shown in Figure \ref{fig:S3_ex_3_1_2} such that $\rm sign (\alpha (\theta_i)) = \rm sign (\alpha)$ and $|\alpha (\theta_i)| < |\alpha|$.
\hfill{\large$\diamond$}
\end{exam}

%

%Note in (b) that the saturation limits are achieved repeatedly for the same values of $y_{\rm df, k}$.

\section{Time-Delayed Lur'e Model with Bias Generation}

%NOTE:  $G(1)\ne0$ in this section

We now modify the discrete-time time-delay Lur'e model by including the bias-generation mechanism shown in Figure \ref{CT_TDL_offset_blk_diag}.
The corresponding closed-loop dynamics are thus given by
\begin{align}
     \left[ \arraycolsep=1.1pt\def\arraystretch{1.2} \begin{array}{c} x_{k+1}\\ x_{\rmd,k+1}\\ x_{\rmf,k+1} \end{array}\right]
     = \left[ \arraycolsep=1.6pt\def\arraystretch{1.2} \begin{array}{ccc} A & 0 & 0\\ 
    e_{d,d}C & N_{d} & 0\\ 0 & e_{1,d}^\rmT   & 0 \end{array}\right] 
     \left[ \arraycolsep=1.1pt\def\arraystretch{1.2} \begin{array}{c} x_{k}\\ x_{\rmd,k}\\ x_{\rmf,k} \end{array}\right]
    + \left[ \arraycolsep=1.1pt\def\arraystretch{1.2} \begin{array}{c} B\\ 0 \\ 0 \end{array}\right]  v_{\rmb,k},
    \label{TDLOffeqn}
\end{align}
with $y_k,$ $y_{\rmd,k},$ and $y_{\rmf,k}$  given by \eqref{TDLeqnlin}, \eqref{ydk}, and \eqref{yfk}, respectively, where $\beta$ is a constant,
\begin{equation}
    v_{{\rm b},k} = (\beta + v_{{\rm f},k})v_k,
\end{equation}
and $v_{{\rm f},k} = {\rm sat}_\delta (y_{{\rm f},k}).$
Note that the constant $\alpha$ is now omitted.
Instead, the constant input $v$ is injected multiplicatively inside the loop, thus playing the role of $\alpha$.
This feature allows the offset of the oscillation to depend on the external input.
The resulting bias $\bar{y}$ of the periodic response is thus given by
\begin{equation}
   \bar{y} = v \beta G(1). 
\end{equation}

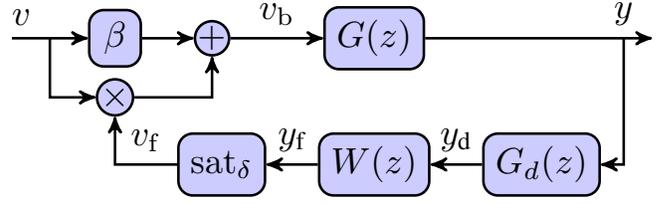
\begin{figure}[h!]
    \centering
    \resizebox{\columnwidth}{!}{%
    \begin{tikzpicture}[>={stealth'}, line width = 0.25mm]
    \node [input, name=input] {};
    \node [smallblock, rounded corners, right of=input, minimum height=0.5cm, minimum width=0.5cm] (beta) {$\beta$};
    \node [sum, right = 0.5 cm of beta] (sum1) {};
    \node[draw = white] at (sum1.center) {$+$};
    \node [smallblock, rounded corners, right = 0.9cm of sum1, minimum height = 0.6cm, minimum width = 0.8cm] (system) {$G(z)$};
    
    \draw [->] (sum1) -- node[name=usys, above] {$v_\rmb$} (system);
    \node [output, right = 2.2cm of system] (output) {};
    \node [smallblock, rounded corners, below = 0.6cm of system, minimum height = 0.6cm, minimum width = 0.8cm](diff){$W(z)$};
    \node [smallblock, rounded corners, right = 0.5cm of diff, minimum height = 0.6cm, minimum width = 0.8cm] (delay) {$G_d(z)$};
    %\node [saturation block, left = 0.8cm of diff, minimum width=1.25cm, minimum height=2.5em] (satq) {};
    \node [smallblock, rounded corners, left = 0.5cm of diff, minimum height = 0.6cm, minimum width = 0.8cm](satq){${\rm sat}_\delta$};
    \node [mult, below = 0.1cm of beta, minimum size=0.35cm] (mult1) {};
    \node [draw = white] at (mult1.center) {$\times$};
    
    \draw [draw,->] (input) -- node [name=u]{} node [very near start, above] {$v$} (beta);
    \draw [->] (u.center) |- (mult1);
    %\draw [->] (beta) -- node [above] {$+$} (sum1);
    %\draw [->] (mult1) -| node [very near end, xshift = -0.25cm] {$+$} (sum1);
    \draw [->] (beta) -- (sum1);
    \draw [->] (mult1) -| (sum1);
    \draw [->] (satq) -|  
    node [near start, above] {$v_\rmf$} (mult1);
    \draw [->] (system) -- node [name=y, very near end]{} node [very near end, above] {$y$}(output);
    \draw [->] (y.center) |- (delay);
    \draw [->] (delay) -- node [above] {$y_\text{d}$} (diff);
    \draw [->] (diff) -- node [above] {$y_{\text{f}}$}(satq);
    \end{tikzpicture}
    }
    \caption{\footnotesize Discrete-time time-delayed Lur'e  model with constant input $v$ and bias generation.}
    \label{DT_TDL_offset_blk_diag}
\end{figure}

\begin{figure}[h!]
    \centering
    \includegraphics[width=0.92\columnwidth]{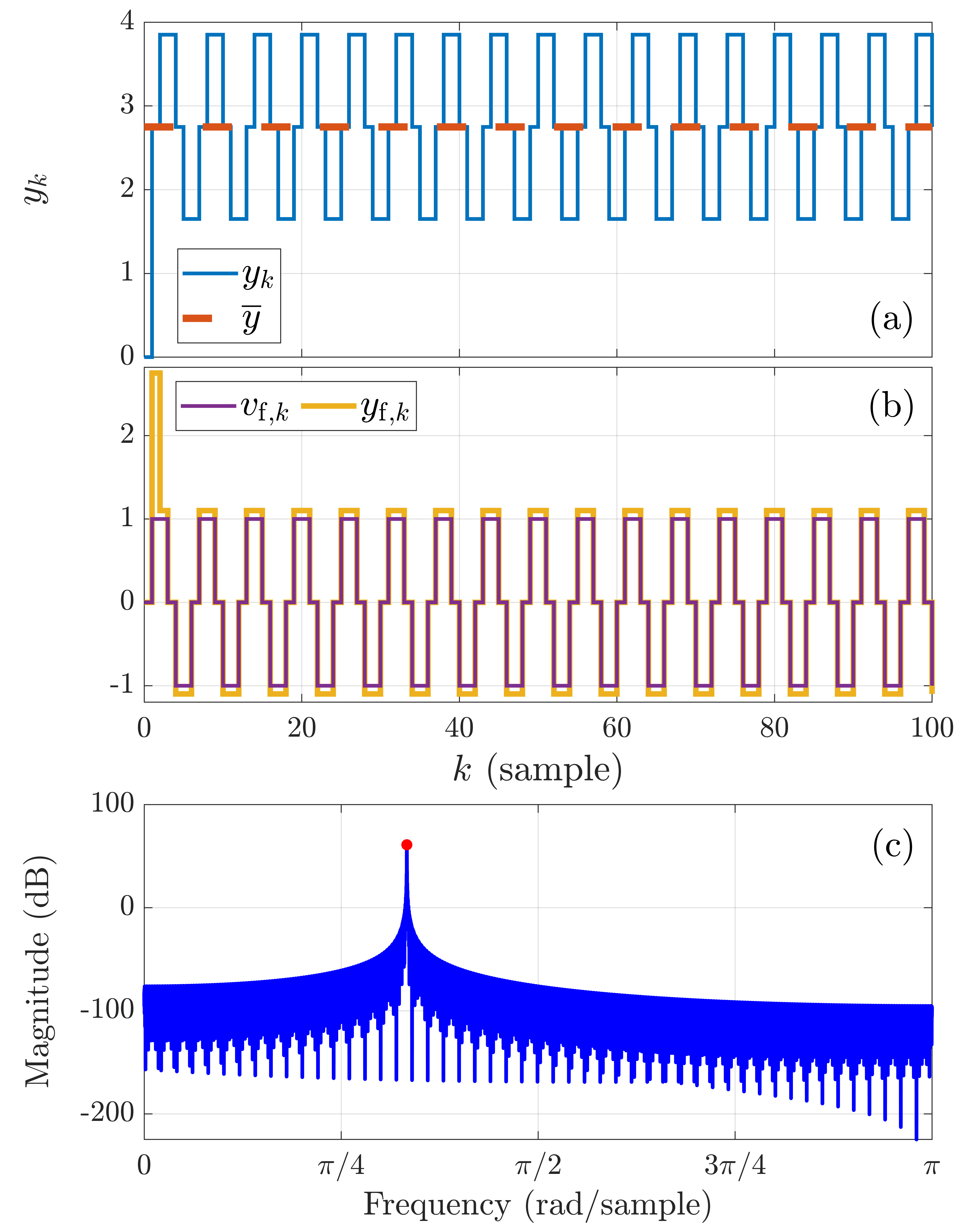}
    \caption{\footnotesize Example \ref{ex_4_1}: For $v = 1.1,$ $\beta = 2.5,$ $d = 0,$ and $\delta = 1,$ (a) shows $y_k$ and the offset $\bar{y}$, (b) shows $v_{{\rm f}, k}$  and $y_{{\rm f}, k},$ and (c) shows the frequency response of $y_k - \bar{y}.$ }
    \label{fig:S4_ex_4_1_1}
\end{figure}

%\textbf{Example 4.1.}
\begin{exam}
\label{ex_4_1}
Let $G(z)\mspace{-2mu}  =\mspace{-2mu} 1/z,$ 
%
%
%
%no, we do not want to say "limit cycle" since we are NOT showing existence of a limit cycle.  Please note the difference between an oscillatory response and a limit cycle-----I think I explained it in the introduction.  We can discuss it when I get back.  Thanks
%
%An offset will be added to the limit cycle displayed in Example \ref{ex_3_1} for 
%
$d\mspace{-2mu} =\mspace{-2mu} 0,$ $\beta = 2.5,$ $v\mspace{-2mu} =\mspace{-2mu} 1.1,$ and $\delta\mspace{-2mu} =\mspace{-2mu} 1.$ 
%
%Figure \ref{fig:S4_ex_4_1_1} shows the response of the model.
%
Figure \ref{fig:S4_ex_4_1_1}(a) shows that the output $y_k$ is oscillatory with offset $\bar{y} = v \beta G(1) = 2.75.$
Figure \ref{fig:S4_ex_4_1_1}(b) shows $v_{\rmf,k}$ and $y_{\rmf,k}.$
%
%Note that, both the linear and saturated regions of the saturation nonlinearity are activated as in Example \ref{ex_3_1}.
%Note that, the values of $v_{\rmf, k}$ alternate between the linear and saturated regions of the saturation nonlinearity are activated as in Example \ref{ex_3_1}.
%
Note that, as in Example \ref{ex_3_1}, despite the offset $\bar{y}$ of $y_k$, the signals $y_{\rmf, k}$ and $v_{\rmf, k}$ oscillate without an offset.
Finally, Figure \ref{fig:S4_ex_4_1_1}(c) shows the magnitude of the frequency response for $y_k - \bar{y}$. Note that the peak is located near the same frequency as in Example \ref{ex_3_1}, and thus the oscillation frequency remains the same with the addition of the bias-generation mechanism.
\hfill{\large$\diamond$}
\end{exam}

%%\clearpage

\begin{figure}[h!]
    \centering
    \includegraphics[width=0.92\columnwidth]{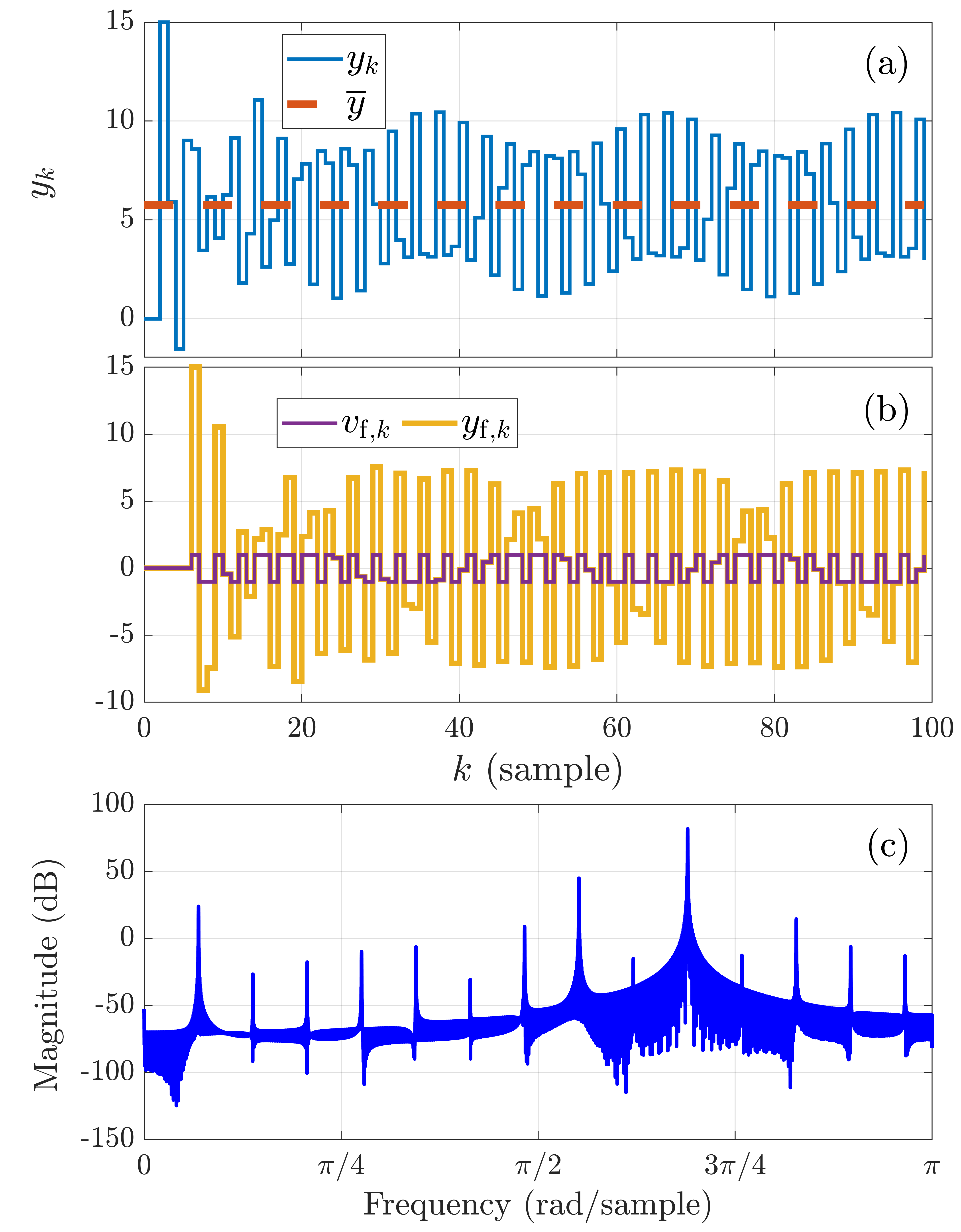}
    \caption{\footnotesize Example \ref{ex_4_2}: For $v = 1,$ $\beta = 15,$ $d = 4,$ and $\delta = 1,$ (a) shows $y_k$ and the offset $\bar{y}$, (b) shows $v_{{\rm f}, k}$  and $y_{{\rm f}, k},$ and (c) shows the frequency response of $y_k - \bar{y}.$ }
    \label{fig:S4_ex_4_2_1}
\end{figure}

\begin{exam}
\label{ex_4_2}
Let
\begin{equation*}
    G(z) = \frac{z - 0.75 e^{\pm \jmath 5 \pi/6}}{(z - 0.9 e^{\pm \jmath \pi/6})(z - 0.9 e^{\pm \jmath 5 \pi/12})},
\end{equation*}
$d = 4,$ $\beta = 15$ $v = 1,$ and $\delta = 1.$ 
Figure \ref{fig:S4_ex_4_2_1}(a) shows that the output $y_k$ is oscillatory with offset $\bar{y} = v \beta G(1) = 5.751.$
%
%Figure \ref{fig:S4_ex_4_2_1}(b) shows that $y_k$ oscillates between the stable and not asymptotically stable regions of the saturation nonlinearity, as in previous cases.
%Figure \ref{fig:S4_ex_4_2_1}(b) shows that the values of $v_{\rmf, k}$ alternate between the stable and not asymptotically stable regions of the saturation nonlinearity, as in previous cases.
%
Figure \ref{fig:S4_ex_4_2_1}(b) shows that $v_{\rmf,k}$ and $y_{\rmf,k}$ have an oscillatory response without an offset, as in previous cases.
Finally, Figure \ref{fig:S4_ex_4_2_1}(c) shows the magnitude of the frequency response for $y_k - \bar{y}$.
\hfill{\large$\diamond$}
\end{exam}

\section{Conclusions and Future Extensions}

This paper presented and analyzed a discrete-time Lur'e model that exhibits self-excited oscillations.
This model involves an asymptotically stable linear system, a time delay, a washout filter, and a saturation nonlinearity.
It was shown that, for sufficiently large loop gains, the response converges to a periodic signal, and thus the system has self-excited oscillations.
A bias-generation mechanism provides an input-dependent oscillation offset.
The amplitude and spectral content of the oscillation were analyzed in terms of the  components of the model.

An immediate extension of this work is to consider the case where $G$ has zeros on the unit circle.
The main results of this paper appear to be valid for this case, although the proofs are more intricate.
Extension to sigmoidal nonlinearities such as atan and tanh as well as relay nonlinearities is of interest.
In addition, continuous-time, time-delay Lur'e models described by {\it iv}) in Section I  are  of interest.
Finally, future work will use this discrete-time self-excited model for system identification and adaptive stabilization.

\section{Acknowledgments}

This research was supported by NSF grant CMMI 1634709,
``A Diagnostic Modeling Methodology for
Dual Retrospective Cost Adaptive Control of Complex Systems.''

%\bibliographystyle{IEEEtran}
%\bibliography{IEEEabrv,bib_paper}

%\clearpage

\section*{Author Biographies}

\noindent {\bf Juan Paredes} received the B.Sc. degree in mechatronics engineering from the Pontifical Catholic University of Peru and a M.Sc. degree in aerospace engineering from the University of Michigan in Ann Arbor, MI. He is currently a PhD candidate in the Aerospace Engineering Department at the University of Michigan.  His interests are in autonomous flight control and control of combustion.

 \medskip

% \begin{wrapfigure}[5]{L}{0.18\textwidth}
%     \vspace{-4ex}
%   \includegraphics[width=.17\textwidth]{aseem.jpg}
% \end{wrapfigure}
%
\noindent {\bf Syed Aseem Ul Islam} received the B.Sc. degree in aerospace engineering from the Institute of Space Technology, Islamabad and is currently pursuing the Ph.D. degree in flight dynamics and control from the University of Michigan in Ann Arbor.
His interests are in data-driven adaptive control for aerospace applications.
 \medskip

%  \begin{wrapfigure}[6]{L}{0.18\textwidth}
%   \vspace{-4ex}
%   \includegraphics[width=.17\textwidth]{AuthorImagesforPaper/Kouba.jpg}
% \end{wrapfigure}
%
\noindent {\bf Omran Kouba} received the Sc.B. degree in Pure Mathematics from the  University of Paris XI and the Ph.D. degree in Functional Analysis from Pierre and Marie Curie University in Paris, France. Currently he is a professor in the Department of Mathematics in the Higher Institute of Applied Sciences and Technology, Damascus (Syria).  His interests are in real and complex analysis, inequalities, and problem solving.

 \medskip

% \begin{wrapfigure}[6]{L}{0.18\textwidth}
%   \vspace{-4ex}
%   \includegraphics[width=.17\textwidth]{dennis.jpg}
% \end{wrapfigure}
%
\noindent {\bf Dennis S. Bernstein} received the Sc.B. degree from Brown University and the Ph.D. degree from the University of Michigan in Ann Arbor, Michigan, where he is currently professor in the Aerospace Engineering Department.  His interests are in identification, estimation, and control for aerospace applications.  He is the author of {\it Scalar, Vector, and Matrix Mathematics}, published by Princeton University Press.

\end{document}